\documentclass[9pt,twocolumn,twoside]{pnas-new-arxiv}

\usepackage[capitalize]{cleveref}
\usepackage{siunitx}
\DeclareSIUnit\Molar{\textsc{m}}

\usepackage{comment}

\templatetype{pnasresearcharticle}

\setboolean{displaywatermark}{false}

\newcommand{\pdt}[1]{\partial_t {#1}}

\renewcommand{\v}[1]{\ensuremath{\mathbf{#1}}} 
\newcommand{\gv}[1]{\ensuremath{\mbox{\boldmath$ #1 $}}} 
\newcommand{\diff}{\mathrm{d}}
\renewcommand{\d}[2]{\frac{\diff #1}{\diff #2}} 

\newcommand{\grad}[1]{\gv{\nabla} #1} 
\renewcommand{\div}[1]{\gv{\nabla} \cdot #1} 
\newcommand{\mean}[1]{\left< #1 \right>}

\newcommand{\norm}[1]{\left\Vert #1 \right\Vert}

\newcommand{\Capillary}{\mathrm{Ca}}

\newcommand{\Bond}{\mathrm{Bo}}
\newcommand{\Peclet}{\mathrm{Pe}}
\newcommand{\dinl}[2]{\mathrm{d} #1/\mathrm{d} #2}
\title{Dynamic multiphase flow triggers chaotic mixing in porous media}

\author[a,b,1]{Gaute Linga}
\author[a]{Kevin Pierce}
\author[a]{Marcel Moura}
\author[c,d]{Joachim Mathiesen}
\author[c,e]{Fran\c{c}ois Renard}
\author[c,f]{Tanguy Le Borgne}

\affil[a]{PoreLab, The Njord Centre, Department of Physics, 0316 Oslo, University of Oslo, Norway}
\affil[b]{PoreLab, Department of Physics, Norwegian University of Science and Technology, Norway}
\affil[c]{The Njord Centre, Departments of Geosciences and Physics, 0316 Oslo, University of Oslo, Norway}
\affil[d]{Niels Bohr Institute, University of Copenhagen, 2100 København Ø, Denmark}
\affil[e]{Univ. Grenoble Alpes, Univ. Savoie Mont Blanc, CNRS, IRD, Univ. Gustave Eiffel, ISTerre, 38000 Grenoble, France}
\affil[f]{Géosciences Rennes, Université de Rennes, CNRS, Unité Mixte de Recherche 6118, 35000 Rennes, France}

\leadauthor{Linga}

\significancestatement{
Chemical and biological processes in soils, rocks, and industrial porous media are often controlled by the mixing of chemicals by fluid flow.
When a single fluid phase carries dissolved chemicals through a porous medium, the mixing process is fairly well understood. 
However, mixing when two or more fluids move together has remained largely unexplored, despite its widespread importance.
We conducted experiments and extensive computer simulations to investigate this process. 
Our findings reveal that moving fluid-fluid interfaces can generate chaotic mixing, characterized by solute blobs stretching exponentially over time.
Hence, mixing in unsteady two-phase flows is qualitatively different and much more efficient than in steady single or two-phase flows, potentially leading to faster mixing and chemical reactions than previously thought.
}
\authorcontributions{
Author contributions: G.L.\ and T.L.B.\ designed research; G.L.\ and J.M.\ developed numerical methods and carried out simulations, K.P.\ and M.M. designed experiments, K.P.\ performed experiments, G.L., K.P., F.R., J.M., and T.L.B.\ analyzed data; G.L., K.P., M.M, F.R., J.M., and T.L.B. wrote the paper.}
\authordeclaration{The authors declare no competing interests.}
\correspondingauthor{\textsuperscript{1}To whom correspondence should be addressed. E-mail: gaute.linga@ntnu.no}

\keywords{chaotic mixing $|$ multiphase flow $|$ porous media} 

\begin{abstract}
Solute mixing plays a pivotal role in a broad spectrum of chemical and biological processes across natural and engineered porous media.
However, current understanding of mixing dynamics remains largely constrained to steady flows in fully or partially water-saturated environments.
Multiphase flow systems are generally unsteady, with moving fluid interfaces and flow paths that change in time.
Despite the widespread occurrence of dynamic multiphase flows, their impacts on solute mixing are largely unknown.
Here, we use experiments and numerical simulations to investigate the effect of dynamic two-phase flow on the stretching and folding of fluid elements, a fundamental mechanism driving solute mixing and reactions in porous media.
We find that dynamic two-phase flows induce chaotic mixing, characterized by exponential stretching of fluid elements, leading to strongly enhanced mixing compared to steady single phase flows.
By extensive numerical multiphase flow simulations, we establish dynamic steady states where we reliably measure the mean fluid stretching rate as a function of flow rate.
We show that stretching is maximized at an optimum flow rate which balances fluid shear deformation against the frequency of flow reorientation by the intermittent motion of the fluid interface.
The findings are rationalized by a mechanistic model linking basic multiphase flow characteristics to the stretching rate, opening new perspectives to understand and control mixing and reactions in a wide range of multiphase flow systems.
\end{abstract}

\dates{\today}

\begin{document}

\maketitle

\thispagestyle{firststyle}
\ifthenelse{\boolean{shortarticle}}{\ifthenelse{\boolean{singlecolumn}}{\abscontentformatted}{\abscontent}}{}

\firstpage[10]{3}

\noindent
Mixing of solutes in porous media flows is essential for a wide range of natural and industrial processes \cite{dentz2011mixing, leborgne2025deformation}.
As mixing brings chemical reactants into contact in fluids \cite{villermaux2019mixing}, it plays a central role in a wide range of biogeochemical reactions \cite{valocchi2019mixing}. 
Limited mixing rates in porous media can reduce effective reaction rates by orders of magnitude compared to batch experiments \cite{meile2006scale,aquino2023fluid}.
Under steady and fully fluid-saturated conditions, modeling frameworks of solute spreading and mixing in porous media flows are fairly well established \cite{berkowitz2006modeling,leborgne2013stretching,leborgne2015lamellar, heyman2020stretching, dentz2023mixing}.

However, a large class of flows occur under unsaturated conditions, where two or more immiscible fluid phases, such as water and air, coexist in the pore space \cite{blunt2017multiphase,feder2022physics,degennes1983hydrodynamic}.
Moreover, many natural and industrial processes exhibit dynamic multiphase flow, where different fluids displace each other, and evolving fluid interfaces continuously reshape the flow pathways over time \cite{reynolds2017dynamic}.
Significant advances have been made in our understanding of solute transport under steady unsaturated conditions, i.e., for water flow past trapped air clusters \cite{degennes1983hydrodynamic,hasan2020direct,velasquez2022sharp}, and multiphase flow has been shown to  enhance solute spreading \cite{mathiesen2023dynamic}.
Experimental observations \cite{tekseth2024multiscale,bultreys20244d} have revealed that sudden fluid rearrangements (e.g.,\ Haines jumps \cite{haines1930studies,berg2013real,tekseth2024multiscale}) can produce significant flow perturbations extending over 20 pore sizes \cite{bultreys20244d}.
However, it is not known to what extent such dynamic multiphase flows impact mixing.

Here, we address this question by resolving experimentally and numerically fluid stretching, the advective driver of mixing, in  multiphase flows through archetypal porous media consisting of randomly arranged cylindrical (2D) or spherical (3D) obstacles.
Our findings reveal that chaotic advection, and thus mixing, can be strongly enhanced by dynamic multiphase flow. 
We show that the magnitude of chaotic mixing is controlled by the capillary number $\Capillary = \mu U / \sigma$, where $\mu$ is the dynamic viscosity of the wetting fluid, $U$ is the mean flow velocity, and $\sigma$ is the interfacial tension.
We identify fluid-interface motion as a major driver of chaotic mixing dynamics, and, importantly, we discern that the competition between shear deformation in moving fluid clusters, which generally decays with $\Capillary$, and reorientation by random interface motion, which becomes more frequent with increasing $\Capillary$, leads to an optimal $\Capillary$ for stretching, with potential implications for mixing processes in a wide range of unsaturated natural and industrial porous media.
\section*{Results}
\subsection*{Experimental and Numerical Evidence of Chaotic Mixing Driven by Multiphase Flow}
We experimentally characterize the qualitative differences in solute mixing between steady single-phase flow and multiphase flow using fluorescence imaging methods.
As it is a well-studied canonical system for single-phase flow, we consider a 3D-printed quasi-2D porous cell consisting of cylinders of diameter $d=\SI{2}{mm}$ packed according to the random sequential adsorption (RSA) algorithm \cite{feder1980random}.
In all cases, we nondimensionalize time by the advective time $t_a = d/U$, with $U$ the average fluid velocity. The capillary number for this experiment is approximately $\Capillary \simeq 10^{-4}$.
The model preparation and experimental techniques are detailed in \textit{Materials and Methods}.

\begin{figure}[htb]
    \centering
    \includegraphics[width=0.99\linewidth]{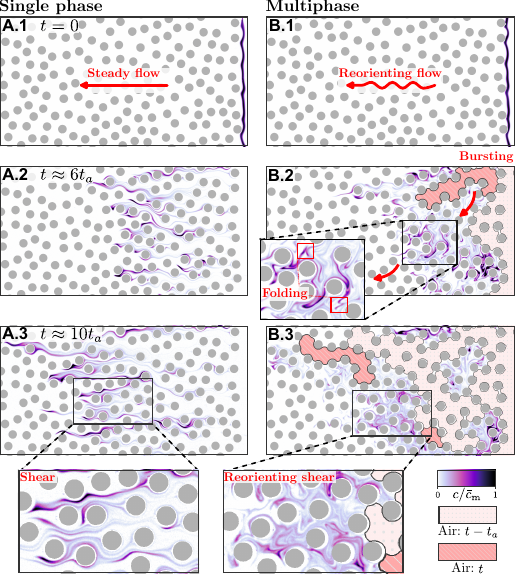}
    \caption{Experimental evidence of chaotic mixing in multiphase flow and comparison to established mechanisms of steady-single phase flow under analogous conditions. Time evolution is displayed in terms of advective times $t_a$. (A) Direct imaging of solute mixing in steady single-phase flow (Movie S1), where the solute front undergoes shear and algebraic stretching, resulting in slow, purely shear-driven mixing at the pore scale, as evidenced in the inset. (B) Direct imaging of solute mixing in a two-phase drainage flow (Movie S2). Interfacial bursts repeatedly reorient the flow direction and fold the solute distribution (B.2 and inset). When alternated with shear, the reorientation sequence produces multiplicative stretching and enhances mixing. 
    Red colors depict interface motion between $t-t_a$ and $t$.
    The multiplicative stretching produces a more stretched and homogeneous concentration distribution than for single-phase flow (compare insets B.3 and A.3).}
    \label{fig:experiment}
\end{figure}

\begin{figure*}
    \centering
    \begin{tikzpicture}[scale=1.0]
    \def\tagh{2.3cm}
    
    \node[] (A) at (0,0) {\includegraphics[width=0.2\textwidth]{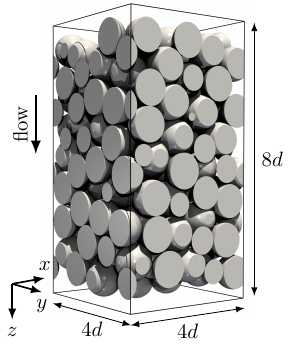}};
    \node[] (Atag) at (-1.4cm, \tagh) {\textsf{\textbf{A}}};

    \node[] (B) at (5.5cm,0) {\includegraphics[width=0.4\textwidth,angle=0,origin=c]{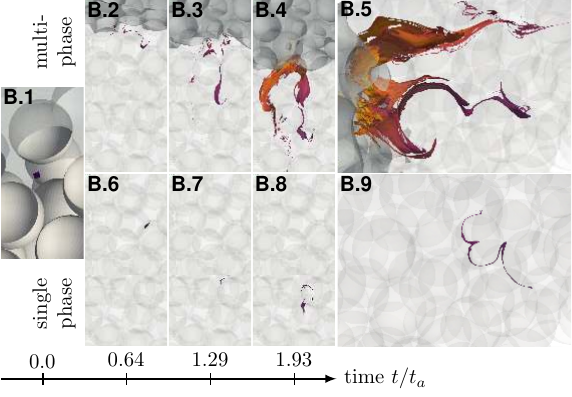}};
    \node[] (Btag) at (2cm, \tagh) {\textsf{\textbf{B}}};
    
    \node[] (C) at (12.4cm,0) {\includegraphics[width=0.35\textwidth,trim=0.4cm 0.4cm 0.4cm 0.2cm,clip]{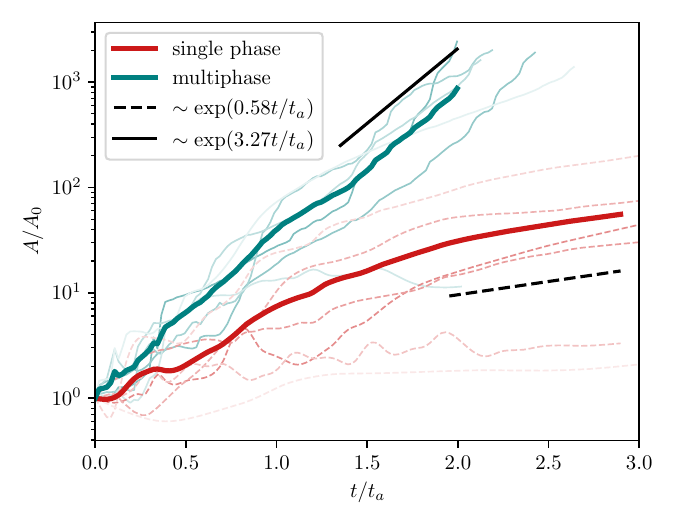}};
    \node[] (Ctag) at (9.5cm, \tagh) {\textsf{\textbf{C}}};
    \end{tikzpicture}
    \caption{Dynamic two-phase flow enhances growth of fluid interfaces compared to single-phase flow. 
    (A) Simulation geometry of granular porous media. The geometry is a periodic packing of identical grains of diameter $d$ constructed using a discrete element method (see \textit{Materials and Methods}). 
    (B) Numerically simulated growth of a Lagrangian sheet with identical initial condition (B.1) in multiphase (B.2-B.4, Movie S4) and single-phase flow (B.6-B.8, Movie S3). (B.5) shows a rotated close-up of the deformed sheet in (B.4), and (B.9) shows the same scene as a rotated close-up of (B.8). The figure reveals that multiphase stretching is clearly more efficient under these conditions.
    (C) Surface area $A$ of sheets relative to their initial size $A_0$ for a two-phase (green solid lines) and a single-phase flow (red dashed lines).
    The flow simulations were performed in the same geometry and with the same flow rate ($\Capillary \simeq 10^{-4}$).
    The thick solid lines represent averages over six individual realizations (thin lines) with identical initial conditions in the two flows, leading to very different trajectories.
    Trajectories were truncated when a point on the sheet first exited the domain.
    The solid black line was fitted using the last 1/3 of the multiphase data.
    The dashed black line is based on established single-phase estimates \cite{heyman2020stretching}.
    } 
    \label{fig:3dsim}
\end{figure*}
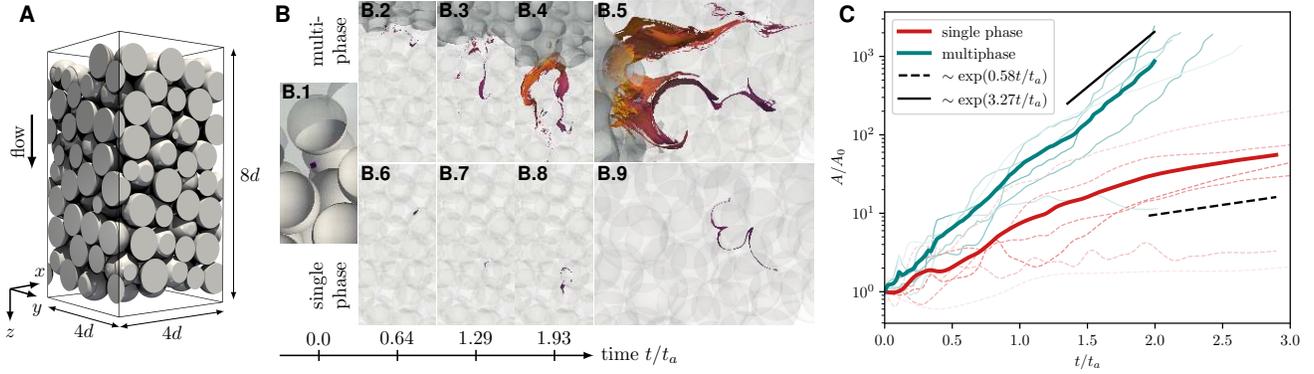

In~\cref{fig:experiment}, we compare steady single-phase flow to drainage flow under matched experimental conditions to demonstrate qualitative differences in solute mixing.
In both cases, we image the concentration field of a fluorescent dye evolving from an initial line distributed transversely to the mean flow.
The steady single-phase flow experiment (\cref{fig:experiment}A and Movie S1) shows that the tracer line deforms into stretched filaments mostly aligned with the flow direction and without reoriented folding patterns (inset of \cref{fig:experiment}A.3). This observation is consistent with the algebraic fluid stretching observed in previous experiments in saturated 2D porous media \cite{borgman2023solute}.
The multiphase flow experiment, where air displaces the wetting liquid, is shown in \cref{fig:experiment}B, displaying folded and densely-packed filaments, suggestive of chaotic mixing \cite{heyman2021scalar}.
Here, the air-liquid front moves downstream in a burst sequence (see \cref{fig:experiment}, panels B.2 and B.3 and Movie S2), with locally high velocities near pore-filling events, i.e.,\ Haines jumps or bursts \cite{bultreys20244d,tekseth2024multiscale}.
These bursts alter the flow field around the interface, reorienting the flow from the mean direction, folding the solute distribution, and ultimately increasing the attack surface for shear (inset of \cref{fig:experiment}B.2).
Across multiple bursts, alternation between reorientation and shear provides a multiplicative stretching mechanism.
The latter is known as the hallmark of chaotic advection, as it produces exponential stretching, persistent concentration gradients, and faster mixing than can be expected from shear alone \cite{villermaux2019mixing}.
These experimental results demonstrate that dynamic two-phase flow has a profound impact on both solute stretching and mixing in prototypical quasi-2D geometries, revealing a sharp transition from shear deformation to chaotic advection. How this behavior translates to 3D geometries is not obvious, since mixing in 3D is inherently chaotic even under single-phase flow conditions due to the additional degree of freedom \cite{heyman2020stretching}.
Measuring fluid stretching at the pore scale in unsteady multiphase flows through 3D porous media remains beyond the reach of current experimental methods.
To circumvent these experimental limitations, we use direct numerical simulations and Lagrangian particle tracking to quantify the stretching of sheets (sheets in 3D are analogous to lines in 2D) and to investigate the differences between single-phase and multiphase flow under otherwise identical conditions. Our simulation procedure is detailed in \textit{Materials and Methods}.
The porous bead-pack geometry is shown in \Cref{fig:3dsim}A as obtained from discrete element simulations of identical beads with diameter $d$, see \textit{Materials and Methods}.
\Cref{fig:3dsim}B shows the qualitative difference in how a small Lagrangian sheet with identical initial conditions evolves in a single-phase (\Cref{fig:3dsim}B.6-9) versus a multiphase flow at $\Capillary \simeq 10^{-4}$ (\Cref{fig:3dsim}B.2-5).
The growth of the sheet is greatly enhanced by the interface motion. This observation is further substantiated in \cref{fig:3dsim}C, which plots the sheet area as a function of time (in advective units) for five different initial conditions (see \textit{SI Appendix}, section B.4, and Fig.\ S1 for the initial conditions).
The single-phase case should asymptotically reach steady exponential growth. 
This behavior can be predicted on the basis of results by Heyman et al.\ \cite{heyman2020stretching}, who measured the topological entropy exponent $\mu \simeq 0.29$, i.e.\ for line growth $\ell \sim e^{\mu t/t_a}$.
For a sheet initially spanned by two perpendicular lines, the growth of the two lines should be uncorrelated, so it is expected that the area of the sheet will grow like $A \sim \ell^2 \sim e^{2\mu t/t_a}$.
The latter result is superimposed in \cref{fig:3dsim}C and compares well with the asymptotic slope of the measurement.
In comparison, the measured exponential growth of the area under multiphase conditions is approximately five times higher.
Although flow perturbations generated by the moving interface, and thus chaotic advection, will likely decrease with distance from the interface, this observation supports the hypothesis that multiphase flow can strongly enhance mixing under realistic flow conditions in porous media.
\subsection*{Optimal Fluid Stretching in Dynamic Multiphase Flows}
In order to extract statistically robust stretching rates, we simulate a dynamic multiphase flow system in a statistically steady regime.
We achieve this by considering a co-flow scenario where two immiscible phases flow simultaneously in a porous medium \cite{avraam1995flow,tallakstad2009steady,datta2014fluid,reynolds2017dynamic,mathiesen2023dynamic}, which we simulate in a large 2D doubly-periodic domain using the finite element method (see \textit{Materials and Methods}).
We initialize the domain with randomly placed cylinders with two matched phases, saturations of the wetting and non-wetting fluids $S_w = S_{nw} = 0.5$, and with a contact angle of \SI {60}{\degree}.
We drive the flow with a uniform average pressure gradient, equivalent to a body force $F$ (per unit volume), until a dynamic steady state is reached. In this regime, fluid clusters are constantly reconfigured, while the total flow rate and saturation fluctuate around a constant value both in space and time. Fluctuations therefore reach an ergodic regime, which facilitates the estimation of robust stretching rates by space and time averaging.
In dimensionless terms, the flow is force-controlled through the dimensionless Bond number $\Bond = F d^2 / \sigma$, where $d$ is the cylinder diameter, $\sigma$ is the surface tension, and $F$ is the external driving force. 

\begin{figure}[htb]
    \centering
    \begin{tikzpicture}
    \def\tagh{1.6}
    \node [] (A) at (0,0) {\includegraphics[width=0.48\linewidth,clip,trim=0.35cm 0.4cm 0.05cm 0.3cm]{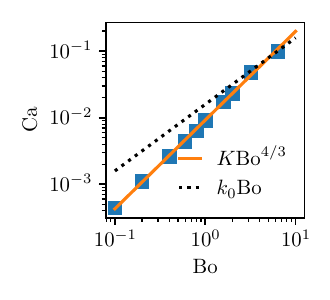}};
    \node [] (Atag) at (-2, \tagh) {\textsf{\textbf{A}}};
    \node [] (B) at (4.2,0) {\includegraphics[width=0.52\linewidth,clip,trim=0.2cm 0.4cm 0.05cm 0.3cm]{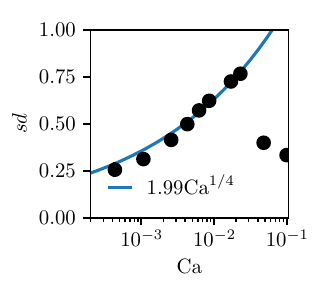}};
    \node [] (Btag) at (2.1, \tagh) {\textsf{\textbf{B}}};
    \end{tikzpicture}
    \caption{Key flow characteristics of statistically steady 2D two-phase flow simulations in a periodic porous medium. The solid symbols represent simulation data. (A) The capillary number, $\Capillary$, ($\propto$ flow rate) is a nonlinear function of the Bond number, $\Bond$ ($\propto$ driving force). The dotted line represents the linear relation obtained from single-phase simulations, where $k_0 \simeq 0.016$ is the dimensionless permeability of the medium. The constant $K = 0.0091$ is determined by a linear best fit to the log-log data.
    (B) specific interface length $s$ as a function of $\Capillary$. This quantity grows with $\Capillary$ until the flow short-circuits, and large clusters span the system. The blue line represents the scaling form for fluid clusters described by \cref{eq:ell_int_scaling}. Since $s$ has the unit of an inverse length, we normalize it by multiplying with the grain size $d$.
    }
    \label{fig:2d_sim_keyflowchar}
\end{figure}

The relationship between the flow rate, characterized by $\Capillary$, and the driving force, quantified by $\Bond$, is generally non-linear due to energy dissipation induced by interfacial forces \cite{tallakstad2009b, mathiesen2023dynamic}.
In \cref{fig:2d_sim_keyflowchar}A, we represent $\Capillary$ as a function of $\Bond$ for a wide range of driving forces. Consistently with previous observations, we obtain the approximate relationship:
\begin{align}
    \Capillary \sim \Bond^{4/3}.
    \label{eq:Ca_vs_Bo}
\end{align}
In the statistically stationary state, neither phase is typically connected throughout the system, and the clusters of both phases continuously split and merge, leading to a dynamic equilibrium between the two processes \cite{talon2023fragmentation,avraam1995flow,tallakstad2009steady,tallakstad2009b,mathiesen2023dynamic}.
To characterize the spatial structure of the fluids in this dynamic equilibrium, we introduce the specific interface length $s = \ell_{\rm int}/ A_{\rm f} $, defined as the interface length ($\ell_{\rm int}$) per fluid area ($A_{\rm f} = \phi L_x L_y$).
This quantity is analogous to the specific interfacial area in 3D \cite{niessner2008model}.
Note that $s$ has units of inverse length, so that $s^{-1}$ is a characteristic length of the phase distribution \cite{debye_scattering_1957, jin_statistics_2016}.
In \cref{fig:2d_sim_keyflowchar}B, we plot $s$ versus $\Capillary$ and identify a generally increasing relationship.
The increase occurs because the cluster size decreases with $\Capillary$ as a result of the balance between capillary and viscous forces \cite{mathiesen2023dynamic}, except at the largest $\Capillary$, where large clusters short-circuit the flow across the periodic boundary (consider the two largest $\Capillary$ values in Fig.\ \ref{fig:2d_sim_keyflowchar}B).
In \textit{Materials and Methods}, we show that
\begin{align}
    s \sim d^{-1} \Capillary^{1/4},
    \label{eq:s_vs_Ca}
\end{align}
which is in good agreement with numerical simulations (Fig. \ref{fig:2d_sim_keyflowchar}B).
For sufficiently low $\Capillary$, the largest cluster of either phase spans the whole system, leading to reduced interface motion, and it is likely that the flow field will eventually become time independent.
On the other hand, for sufficiently large $\Capillary$, the interface forces play a decreasingly small role, and the fast flow creates streamwise channels connected to themselves through the periodic boundary.

\begin{figure*}
    \centering
    \includegraphics[width=0.9\textwidth]{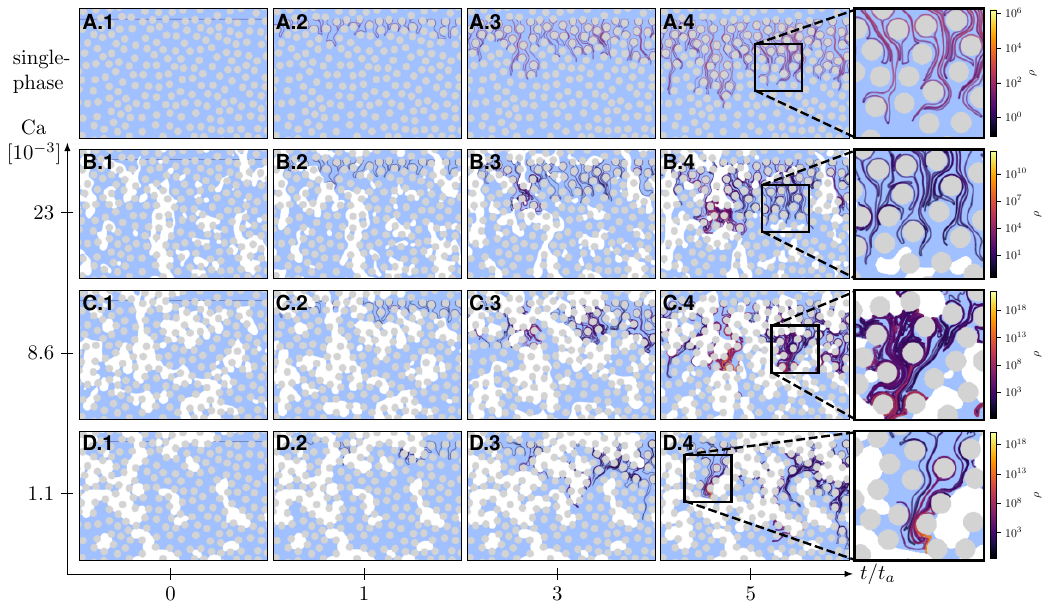}
    \caption{Representative time evolution of stretching and folding in simulated multiphase flow at different $\Capillary$. 
    Stretching appears optimal for $\Capillary \simeq 10^{-2}$.
    The gray circles represent solid obstacles, blue represents the wetting phase, white is the non-wetting phase, and the color scale from black to yellow refers to the local elongation $\rho$ of the solute filament.
    We show the same, small part of a periodic $60d \times 90d$ double-periodic domain in statistically steady states corresponding to different $\Capillary$.
    (A.1-4) Stretching in steady single-phase (i.e.\ fully saturated) flow serves as a baseline and displays pure stretching with no folding.
    (B.1-4) For multiphase flow, at relatively high capillary number $\Capillary = 2.3 \cdot 10^{-2}$, the stretching is similar to that in steady flow, with small-amplitude folding events leading to exponential stretching.
    (C.1-4) Intermediate $\Capillary$ shows sustained stretching and folding in most of the domain.
    (D.1-4) Lower $\Capillary$ shows large stretching and folding events in parts of the domain, and relative stagnation in other parts. See Movies S5-S8 for visualizations of the simulations A-D, respectively.
    }
    \label{fig:line_2d_stretching}
\end{figure*}

We therefore limit the analysis to the range between these two extremes, i.e., $\Capillary \in [10^{-4}, 10^{-1}]$ for the flow conditions considered.
In \cref{fig:line_2d_stretching} we show snapshots over five advective times $t_a$ for three representative values of $\Capillary$ where the flow is in a dynamic steady state.
Zoomed-in snapshots show representative portions of the domain under different $\Capillary$.
In \cref{fig:line_2d_stretching}A we show the steady single-phase (fully-saturated) case as a reference (Movie S5).
At high $\Capillary$ (\cref{fig:line_2d_stretching}B and Movie S6), the non-wetting clusters are elongated in the streamwise direction and move easily, only perturbing the local velocity field in the wetting phase.
At intermediate $\Capillary$ (\cref{fig:line_2d_stretching}C and Movie S7), the non-wetting clusters are larger and the interfaces may get pinned at local pore constrictions, leading to more drastic changes in flow directions. 
Stagnant zones are nevertheless not prevalent, and most of the fluid domain is mobilized.
At the lower $\Capillary$ (\cref{fig:line_2d_stretching}D and Movie S8), the clusters are even larger, and the interface evolves only in a few pores along the interface at a time. 
This intermittent behavior relates to the dynamics of Haines jumps \cite{haines1930studies,berg2013real,tekseth2024multiscale} and can lead to strong velocity fluctuations focused near the interface \cite{bultreys20244d} and abrupt reorientation of local flow directions.

To assess how the different flow regimes described above affect solute stretching, we place a strip of advective particles initially transverse to the mean flow, as shown with a horizontal black line through the wetting phase in \cref{fig:line_2d_stretching}A-D.1, and we numerically advect the material line \cite{meunier2010diffusive,leborgne2015lamellar}, as described in \textit{Materials and Methods}. 
We monitor the local elongation $\rho = \delta \ell(t) / \delta \ell(0)$, where $\delta \ell(t)$ is the infinitesimal separation between two adjacent material points along the line.
As a baseline, the stretching in the single-phase case progresses as expected, appearing qualitatively similar to the single-phase experiment (\cref{fig:experiment}A and Movie S2) with material elements aligning with the direction of flow due to shear and no randomly-reoriented folding (\cref{fig:line_2d_stretching}A and Movie S5).
For the multiphase case, the deformation dynamics are markedly different and depend on $\Capillary$.
For the higher $\Capillary$ case (\cref{fig:line_2d_stretching}B and Movie S6), the line deformation appears similar to steady single-phase flow (\cref{fig:experiment}A and Movie S1), although small-amplitude folding events occur in most of the domain due to cluster rearrangements.
For the intermediate $\Capillary$ case (\cref{fig:line_2d_stretching}C and Movie S7), the line deformation is qualitatively similar to the drainage experiment (\cref{fig:experiment}B and Movie S2), with large-amplitude folding events that occur in most of the domain.
The lower $\Capillary$ case (\cref{fig:line_2d_stretching}D and Movie S8) exhibits similar large-amplitude folding events, but they appear only in localized parts of the fluid domain, while slow-flowing parts deform more like a single-phase flow. 
In particular, the maximally-stretched segments $\rho_{\rm max}$ are similar in the latter two simulations, as shown by the color bar in \cref{fig:line_2d_stretching}, while the higher $\Capillary$ case (\cref{fig:line_2d_stretching}B) exhibits a narrower distribution.

The mean elongation, $\mean{\rho}(t) = \ell(t)/\ell(0)$, where $\ell(t)$ is the instantaneous total length of the line, increases exponentially in time for all multiphase cases (see \textit{SI Appendix}, Fig.\ S2), consistent with the observation of repeated stretching and folding events. 
Notably, the growth rate in the case of intermediate $\Capillary$ ($\sim 10^{-2}$, \cref{fig:line_2d_stretching}C) is higher than the other two, suggesting a non-monotonic dependence on $\Capillary$. In contrast, the single phase simulation exhibits a linear elongation, consistent with expectations for two-dimensional steady flows (Supplementary Figure S2).
To quantify the dependence of stretching on $\Capillary$, we use a more precise deformation metric to probe stretching uniformly in the entire domain over through time.
We measure the instantaneous growth rate as $\tilde \lambda(t) = \mean{\dinl{\ln \rho}{t}}$ where $\rho = |\gv \rho|$ is the elongation, and the average is taken over $10^4$ passive particles that accumulate deformation of their elongation vector $\gv \rho$ according to
\begin{align}
    \d {\gv \rho} t = \gv \rho \cdot \grad \v u, \quad \gv \rho(0) = \cos \theta_0 \hat{\v x} + \sin \theta_0 \hat{\v y},
    \label{eq:drhot}
\end{align}
where $\v u$ is the flow velocity and $\theta_0$ represents the initial angle of the elongation vector.
Initially, the particles are distributed uniformly in space throughout both phases with random initial orientations.
After an initial transient, we find that the instantaneous stretching rate for all $\Capillary$ tend to a constant value $\tilde \lambda(t) \to t_a^{-1} \lambda$ (see \textit{SI Appendix}, Fig.\ S4).
We identify $\lambda$ as the \emph{dimensionless} (infinite-time) Lyapunov exponent, which reflects the mean stretching experienced by the particles throughout the domain, relative to the mean flow (i.e.,\ per advective time $t_a$).

\begin{figure}[htb]
    \centering
    \includegraphics[height=5.2cm]{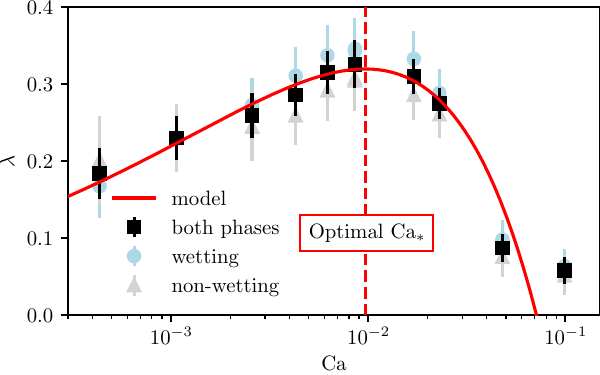}
    \caption{
    Quantitative model predicts stretching optimality in 2D dynamic multiphase porous media flows. Measured Lyapunov exponents exhibit non-monotonic behavior, with a special capillary number $\Capillary_\ast \simeq 10^{-2}$ at which optimal stretching occurs. The reoriented shear model of \Cref{eq:lambda_combined}, plotted as a red line, reveals the competition between reorientation frequency and shear deformation that produces optimal stretching. The figure also shows that filaments in the wetting and non-wetting phases undergo similar stretching.
    }
    \label{fig:steady2dresults}
\end{figure}

In \cref{fig:steady2dresults}, we show the Lyapunov exponents $\lambda(\Capillary)$ obtained for different $\Capillary$ by averaging $\tilde\lambda$ over $t \in I = [30t_a, 60t_a]$.
Consistent with the observations of \cref{fig:line_2d_stretching}, $\lambda(\Capillary)$ exhibits a non-monotonic single-peaked behavior, with a maximum at $\Capillary \simeq 10^{-2}$.
The exponents obtained from particles in the wetting and non-wetting phases are very close and follow the same trend. 
In particular, the maxima of $\lambda, \lambda_w, \lambda_{nw}$ are $0.33 (\pm 0.03)$, $0.35 (\pm 0.04)$, and $0.30 (\pm 0.04)$, respectively, for both phases, the wetting phase, and the non-wetting phase. 
The exponent for the wetting phase is marginally higher, which can be rationalized by the observation that the wetting phase is in contact with more of the solid obstacles, which on average makes it subject to stronger shear than the non-wetting phase.
An exception is seen for low $\Capillary$, where $\lambda_w < \lambda_{nw}$. Under these conditions, clusters of the wetting phase are more likely to be trapped and stagnant, leading to less stretching than in the non-wetting phase. We note however that the confidence intervals of all three exponents overlap.
\subsection*{Linking Stretching Rates to Multiphase Flow Properties}
To explain the non-monotonic relation between the Lyapunov exponent ($\lambda$) and the capillary number ($\Capillary$), we further investigate the basic mechanisms of fluid deformation in this dynamic multiphase flow system. 
In the bulk of a moving cluster, fluid deformation occurs relatively steadily, as in single phase flow (see Movies S6-S8): material lines wrap around grains and elongate by shear due to the gradient of velocity that develops  between the pore center and the grain walls. On the other hand, when located close to cluster boundaries, material lines are rapidly reoriented as they experience strong velocity fluctuations induced by interface motion. 
We thus consider the mechanism of deformation by reoriented shear \cite{khakhar_fluid_1986, hinch1999mixing}, which produces the stretching rate  $\tilde \lambda = f \log \left( {\dot\gamma}/{f} \right)$, where $\dot\gamma$ is a shear rate and $f$ is a reorientation frequency.  

The average shear rate within fluid clusters $\dot\gamma $ can be estimated from the mean velocity gradient at the pore scale $\dot\gamma \sim U/d = a t_a^{-1}$, with $a$ a prefactor of order unity that depends on the pore geometry. To estimate the reorientation frequency, we describe the probability for a material line to be located at a cluster boundary in terms of the length scale
\begin{equation}
    \Delta=s^{-1}=A_{\rm f} /\ell_{\rm int}.
    \label{eq:delta}
\end{equation}
As the reciprocal of the specific interface length, this quantity represents the characteristic separation distance between like phases in the pore space \cite{debye_scattering_1957, jin_statistics_2016}.
The interface moves at the average fluid velocity $U$, and although the interface length remains constant in the considered dynamic steady state, the interface is constantly reshaped as fluid clusters fragment and merge (see movies S6-S8). These events reorganize the interface over a characteristic timescale $t_{\Delta}=\Delta/U$.
We therefore assume that the distance of a material line to the closest interface  is randomly reset over the timescale $t_{\Delta}$. 
After a fragmentation/merging event,  the probability for a material line to be located within a pore size $d$ of the interface is thus $p  \sim  \ell_{\rm int}d/ A_{\rm f}=d/\Delta $.
The characteristic time at which a material line experiences random reorientation events in the vicinity of an interface is therefore $t_r=t_{\Delta}/p\sim t_a (\Delta/d)^2$. 
Hence the reorientation frequency is $f=1/t_r= b t_a^{-1}({d}/{\Delta})^2$, where $b$ is a prefactor of order unity. We thus obtain the non-dimensional stretching rate $\lambda = \tilde \lambda t_a$ resulting from this process as,
\begin{equation}
    \lambda
    = b\frac{d^2}{ \Delta^2} \log \left(a \frac{\Delta^2}{d^2}\right) = b d^2 s^2  \log \left(  \frac{a}{ d^2 s^2}\right),
    \label{eq:lambda_scaling}
\end{equation}
where $a$ and $b$ are prefactors of order unity.
This conceptual model implies that stretching results from a competition between the reorientation frequency $f\sim d^2/\Delta^2$, which decreases with the characteristic interface scale $\Delta$, and the elongation by shear between reorientation events 
${\dot\gamma}/{f} \sim \Delta^2/d^2$, which increases with $\Delta$.
Taking prefactors $a=b=1$, this model predicts that the optimum characteristic interface scale $\Delta_*$ and stretching rate $\lambda_*$ resulting from the balance of reorientation frequency and shear elongation are respectively $\Delta_*/d=e^{1/2} \approx 1.65 $ and $\lambda_*=e^{-1} \approx 0.37$. The latter is close to the largest Lyapunov exponent measured in the simulations (Fig. \ref{fig:steady2dresults}).

We further express the stretching model as a function of capillary number. From equations \eqref{eq:s_vs_Ca}, the specific interface length is $s d \simeq 2\Capillary^{1/4}$, such that \eqref{eq:lambda_scaling} yields:
\begin{align}
    \lambda = b' \Capillary^{1/2} \left[\log a'  -  \log \Capillary^{1/2}\right],
    \label{eq:lambda_combined}
\end{align}
where $b'=4b$ and $a'=a/4$.
We fit the model to the numerical data by adjusting the two prefactors to $a'=0.267$ and $b'=3.25$ respectively. This leads to prefactors for equation \eqref{eq:lambda_scaling} $a=1.06$ and $b=0.82$ of order unity, as expected.
The model is in good agreement with the data (Fig. \ref{fig:steady2dresults}), capturing the non-monotonic dependence of chaotic stretching on $\Capillary$. 
\Cref{eq:lambda_combined} predicts a maximum at $ \Capillary_* = \exp\left(2\log a' -2 \right)$, 
leading to an optimum capillary number $\Capillary_\ast\simeq 10^{-2}$.
The resulting optimum stretching rate is
\begin{align}
     \lambda_* = b' \Capillary_*^{1/2}, 
     \label{eq:lambda_star}
\end{align}
which gives $\lambda_*\simeq 0.34$, in good agreement with the maximal Lyapunov exponent $\lambda_{\rm max} = 0.33 \pm 0.03$ observed in the numerical simulations (see Fig.\ \ref{fig:steady2dresults}). 
%
\section*{Discussion}
\subsection*{Chaotic Advection Triggered by Interfacial Bursts}
This study establishes the role of multiphase flow as a driver for chaotic advection at the pore scale. Using statistically steady multiphase flows in two-dimensional porous media, we developed a mechanistic model that captures the key processes driving chaotic mixing in such systems. Our experimental and numerical results demonstrate that these dynamics are not limited to statistically-steady conditions but also emerge widely in transient flow fronts, both in 2D and 3D porous media (\cref{fig:experiment} and \cref{fig:3dsim}).

Flow reorientation and folding by fluid interface motion is the fundamental mechanism driving chaotic mixing in multiphase flows.
In the experiments (\cref{fig:experiment}), reorientation and folding were driven by interfacial bursts, while in the co-injection simulations (\cref{fig:steady2dresults}), they were caused by the merging and fragmentation of fluid clusters. 
While the intensity of chaotic mixing drops sharply above $\Capillary \sim 10^{-2}$ (\cref{fig:steady2dresults}A), it decays slowly towards low $\Capillary$, suggesting that chaotic mixing is sustained over a broad range of multiphase flow regimes.
The chaotic mixing process described here should play an important role for Péclet numbers $\Peclet \gtrsim 1$ \cite{villermaux2019mixing}, where $\Peclet = U d / D$ and where $D$ is the molecular diffusion coefficient.
We note that $\Peclet$ is directly proportional to $\Capillary = U \mu / \sigma$ through the mean velocity $U$, and, as a consequence, the ratio $\Peclet / \Capillary =  \sigma d / ( \mu D )$ is fixed for a given flow system.
Using typical values for a porous salt-water-air context ($\sigma \sim 10^{-1} \textrm{Nm}^{-1}$, $d \sim 10^{-4} \mathrm{m}$, $D \sim 10^{-9} \mathrm{m^2 s^{-1}}$, $\mu \sim 10^{-3} \mathrm{Pa \, s}$) we find $\Peclet \sim 10^8 \Capillary$.
Hence, the range of capillary numbers where interface-driven chaotic mixing matters can be estimated from the minimum $\Capillary$ where $\Peclet \sim 10$ and a maximum $\Capillary$ where intermittency vanishes (\cref{fig:steady2dresults}A), leading to $\Capillary \in [10^{-7}, 10^{-2}]$.
We thus expect chaotic mixing to persist across a wide range of multiphase flow systems, as the observed mechanisms are inherent to all dynamic two-phase porous flows.

The highest measured values exceed recently-estimated Lyapunov exponents for steady 3D flows through random bead packs, $\lambda \simeq 0.21$ \cite{heyman2020stretching}, ordered bead packs, $\lambda \simeq 0.13$ \cite{turuban2019chaotic}, and pore network models, $\lambda \simeq 0.12$ \cite{lester2013chaotic,lester2016chaotic}.
This observation suggests that the mechanism generating chaotic advection in multiphase flow, i.e.,\ reorientation and folding of fluid elements by sudden interface motions, may dominate over the stretching and folding induced by the 3D pore geometry, as observed in our 3D simulations (\cref{fig:3dsim}).
On the other hand, immobilization of clusters at low $\Capillary$ could potentially limit fluid stretching in multiphase flows.
We note that the Lyapunov exponents found are not expected to be universal for all two-phase flows, and any flow condition that would change the interface area and flow dynamics would likely alter the stretching rate.
\subsection*{Enhanced Mixing and Chemical Reactions in Two-Phase Flow}
Using lamellar mixing models \cite{villermaux2019mixing,leborgne2015lamellar,heyman2020stretching, leborgne2025deformation}, these stretching rates may be coupled with diffusion to obtain estimates of solute mixing at finite Péclet numbers.
Following recent work on mixing in chaotic flows \cite{heyman2020stretching}, we expect the maximum concentration to decay as $c_\textrm{max} \sim \exp(-(\lambda+\sigma_{\log\rho}^2/2)t/t_a)$ in time, where $\sigma_{\log\rho}^2$ is the exponent of the variance of $\log\rho$.
This is in contrast to scalar decay in steady 2D flows, which are expected to decay algebraically, but similar to the expectation in steady 3D porous media flows. 
As shown in the present work, the exponents found in 2D two-phase flow can be larger than those found for steady 3D flow in bead packs and pore networks, which, as reflected in \cref{fig:3dsim}, suggests that chaotic mixing in 3D under steady unsaturated conditions can be dramatically enhanced by interfacial bursts.
In \textit{SI Appendix}, Fig.\ S3, we show that the estimated $c_\textrm{max}$ for the simulations in \cref{fig:line_2d_stretching} averaged along the solute strips, decays exponentially in time for multiphase flow, consistent with exponential scalar decay, while for single-phase flow it decays algebraically.

A key consequence of the resulting enhanced mixing is the potential for accelerating chemical reactions. 
Mixing-induced reactions are critical to a wide range of environmental processes, including remediation of contaminants in the subsurface \cite{kitanidis2012delivery} and trapping of contaminants \cite{datta2009redox}. 
In steady porous media flow, mixing can remain incomplete and limit reactions \cite{sanquer2024microscale}.
The presence of a moving second phase leads to higher mixing rates, and therefore higher reaction rates.
Recent experiments \cite{zhang2025enhanced} found that statistically steady air-water co-flow leads to higher reaction rates and more reaction product created over time than analogous single-phase or unsaturated flows. Our findings hence provide a framework to understand and predict these phenomena.

While multiphase flows can enhance mixing, this effect may be offset by the trapping of clusters that remain behind drainage and imbibition fronts \cite{lenormand1985invasion,birovljev1991gravity,reis2025drainage}. 
If such clusters are stagnant, mixing within them will be purely diffusive.
This effect can inhibit reactions; either by keeping some of the reactant from meeting other reactants or by leaving incompletely mixed reactants together in stagnant clusters.
Thus, for mixing to be efficient in two-phase flow, clusters will have to be efficiently remobilized; otherwise, the effect of chaotic mixing may be temporary. In 3D porous media, the characteristic sizes of the trapped clusters due to drainage are expected to be smaller than in 2D \cite{reis2025drainage}, likely leaving less of the solute trapped.
\subsection*{Optimizing Stretching and Mixing by Two-Phase Flow}
Of particular relevance to microscale mixing is the mixing efficiency $\eta$, i.e.,\ the ratio of the average stretching rate $\tilde \lambda$ to the average strain rate  $\norm{\gv \varepsilon}$ \cite{ottino1990mixing}.
Using the principle that in the steady state, the injected power should be dissipated by viscosity, and so $U F = 2 \mu \norm{\gv \varepsilon}^2$, where we recall that $F$ is the external driving force (macroscopic pressure gradient). We thus obtain:
\begin{align}
    \eta = \frac{\lambda U/d}{\norm{\gv \varepsilon}} = \lambda \sqrt{\frac{2 \mu U}{F d^2}} = \lambda \sqrt{2} \sqrt{\frac{\Capillary}{\Bond}} = \lambda \sqrt{2} K^{3/8} \Capillary^{1/8},
    \label{eq:mixing_efficiency}
\end{align}
where in the last equality we have used \cref{eq:Ca_vs_Bo}.
The mixing efficiency $\eta$ is thus expected to follow the same functional form as \eqref{eq:lambda_combined} with a maximum towards slightly higher $\Capillary$, due to the weak additional factor $\Capillary^{1/8}$. 
Using the maximum value $\lambda_{\rm max} \simeq 0.33$, $\Capillary_* \simeq 10^{-2}$, and the empirically measured $K \simeq 10^{-2}$, we estimate $\eta_{\rm max} \simeq 0.047$.
This value is directly comparable to what is found for industrial mixers \cite{ottino1990mixing}, slightly higher than what was recently found for steady flow through random bead packs \cite{heyman2020stretching}, and an order of magnitude higher than chaotic herringbone mixers \cite{stroock2002chaotic} that are widely used for microscale flows. 
Given the potential to optimize flow conditions beyond what is done in the present study, e.g.,\ by varying surface tension, density and viscosity ratios and saturation, as well as by exploiting 3D flow topologies, multiphase flow-induced chaotic advection could facilitate new and energy-efficient microscale mixing technologies beyond the current state of the art \cite{kreutzer2005multiphase}.
\subsection*{Future Studies and Applications}
The process described here is likely to be present in a wide range of natural and industrial multiphase flow contexts where solute transport and dynamic fluid connectivity occur simultaneously \cite{mathiesen2023dynamic}, such as the flow of hydrogen and water \cite{lysyy2022pore}, gas/CO$_2$ and brine in subsurface reservoirs \cite{reynolds2017dynamic,zheng2017effect}, and solute transport in unsaturated soils \cite{vereecken2022soil}.
Our findings suggest that multiphase flow processes, such as drainage and imbibition, can largely enhance the mixing of solutes in these applications. This can play an important role in mobilizing contaminants \cite{brusseau1994transport}, triggering chemical reactions \cite{valocchi2019mixing} or sustaining microbial activity \cite{bochet2020iron}.

The quasi-2D experiments presented in \cref{fig:experiment} qualitatively confirm the persistence of stretching and folding, but future studies should directly measure concentration decay, complementing recent experiments for reaction rates \cite{zhang2025enhanced}, as well as extend the study to 3D flows relevant to natural settings.
Experimental realizations of mixing in 3D multiphase flow have, until recently, been beyond reach due to the opaque nature of porous media and the high resolutions required to reliably measure solute concentrations, but may now be feasible by combining advancements in imaging mixing \cite{heyman2020stretching,heyman2021scalar,sanquer2024microscale} and two-phase flow \cite{brodin2025interface}.
Recent advances in stroboscopic X-ray microtomography \cite{tekseth2024multiscale} and 4D micro-velocimetry \cite{bultreys20244d} could also enable direct imaging of pore-scale mixing resulting from Haines jumps.
Such studies would be critical for generalizing the observations presented in this study to direct measurements of mixing and reaction that are challenging to assess numerically, covering larger ranges of capillary number, contact angles, Péclet number, Ohnesorge number, viscosity ratio, and pore-scale topology. 
Additional factors that are likely to influence mixing and reaction dynamics include buoyancy, system size, compressibility effects, the trapping of liquid clusters during drainage, and solute transport driven by slow flows in liquid corners and bridges \cite{reis2023simplified,reis2025interaction}.
Building on the principles presented here, the study of coupled multiphase flow and mixing thus opens new opportunities to  understand, model, and control solute transport dynamics in porous media.
\matmethods{
\subsection*{Porous Domains}
In 2D, we consider porous domains generated by random sequential adsorption (RSA), i.e.,\ by depositing as many circular obstacles as possible at random locations in the domain where the circles do not overlap with any already placed obstacles \cite{feder1980random}.
The initial diameter is set to $d_{\rm init}=1.2 d$, and to ensure good connectivity of the pore space, we then shrink the cylinders to a diameter $d$, which produces a porosity of approximately $\phi = 1 - (1-\phi_{\rm RSA}) (d/d_{\rm init})^2 \simeq 0.62$, where $\phi_{\rm RSA} \simeq 0.547$ is the porosity resulting from RSA \cite{feder1980random,zhang2013precise}. 
In the simulations, the RSA algorithm is used with doubly periodic boundary conditions and meshed using the script \texttt{periodic\_porous\_2d.py} implemented in Bernaise \cite{linga2019bernaise}. 
For these simulations, we use a $60d \times 90d$ domain that we found to be large enough to resolve the largest clusters for $\Capillary \gtrsim 10^{-3}$.

For 3D simulations, we consider a random close packing of identical spheres of diameter $d$. 
To prepare the geometry, we specify a target box size $4d \times 4d \times 7d$ and place $N$ spheres in a triple-periodic domain, which we then numerically shrink with fixed aspect ratios, using the discrete element code Yade \cite{yade2024}, until the system is jammed. Depending on the difference between the box size in the jammed state and the target size, we increase or decrease $N$ and repeat the compression process, until the jammed-state box size is within $0.1 \%$ of the target size. The box is then padded by $0.5d$ at both the top and bottom to allow for uniform inlet and outlet velocity fields, such that we end up with a box size $4 \times 4 \times 8 d^3$. The configuration is meshed using the \texttt{mesh-sphere-packing} code \cite{knight2025,knight2020computing}.
\subsection*{Two-Phase Flow Simulations}
We simulate two-phase flow of immiscible matched fluids (same viscosity and density).
In order to obtain a smooth velocity field with negligible spurious currents, which is a common problem for low-$\Capillary$ flows \cite{connington2012review,raeini2012modelling}, we use a diffuse-interface approach based on solving the Navier--Stokes--Cahn--Hilliard model \cite{jacqmin1999calculation,anderson1998diffuse} with a driving force $F$:
\begin{gather}
    \pdt \phi + \div (\v u \phi ) = m \nabla^2 \eta, \quad
    \eta = - \epsilon \nabla^2 \phi + \epsilon^{-1} W'(\phi), \\
    \pdt{ \v u } + \v u \cdot \grad \v u - \nu \grad^2 \v u = - \grad \bar{p} -  \gamma \phi \grad \eta - \frac{F}{\varrho} \hat{\v y}, \quad \div \v u = 0.
\end{gather}
Here $\phi$ is the phase field, which takes the value $1$ in the wetting phase and $-1$ in the non-wetting phase. $\phi$ continuously interpolates between these values across the finite-width interface.
Moreover, $m$ is the mobility, $\eta$ is the phase field chemical potential, $\epsilon$ is the interface thickness, $W = (1-\phi^2)/4$ is a double-well potential, $\nu$ is the kinematic viscosity of both fluids, $\tilde p$ is a reduced pressure, $\gamma = \sigma/\varrho$ is a reduced interfacial tension and $\varrho$ is the density.
The interface is initialized at a distance $0.8 d$ from the inlet.
We use a pressure-splitting scheme for the temporal discretization with $P_1$ finite elements for both velocity components and pressure defined on the triangulated domains described above (in 2D and 3D).
The model is solved using our open-source solver Twoasis \cite{linga2025twoasis} based on Oasis \cite{mortensen2015oasis} and Fenics/Dolfin 2019.2.0 \cite{logg2012automated} on distributed-memory high-performance computing systems (see \textit{SI Appendix}, section A).

The 2D simulations are run in a double-periodic domain and initialized with $\v u = \v 0$ and $\phi$ in a checkerboard pattern of $-1$/$1$ with a length scale $5d$, which results in a saturation of close to $50 \%$. After an initial transient, the system approaches a statistically stationary state where the space-averaged velocity oscillates around a well-defined temporal mean for each given global forcing $F$ \cite{mathiesen2023dynamic} (force controlled conditions). Here the contact angle is fixed at $\theta = \SI{60}{\degree}$, $\epsilon = 0.05$, $\nu = 1$, $\gamma = 5$, $m = 0.0005$.
The 3D simulations are initialized with the interface at a distance $d$ from the inlet. A uniform velocity $\v u = u_0 \hat{\v z}$, with $u_0 = 10^{-3}$ is imposed at both inlet and outlet, representing flux-controlled conditions. We impose a neutral contact angle $\theta = \SI{90}{\degree}$, $\gamma = 20$, $m = 0.002$, $\nu=1$. In both cases, we use a linear element size of $\Delta x \sim 0.028 d$ with $d=1$. The velocity fields are sampled at time intervals $\tau \propto \Delta x / U$ and stored as XDMF/HDF5 files for use with Lagrangian particle tracking.
\subsection*{Measuring Deformation and Stretching}
We measure solute deformation by using Lagrangian particle tracking, in particular by discretizing solute strips/sheets into a discrete set of points $\v x$ that obey:
\begin{align}
    \d {\v x} t = \v u (\v x(t), t),
    \label{eq:dxt}
\end{align}
where $\v u$ is the numerically obtained velocity field.
The points are connected by edges or triangles, which are refined or coarsened as the structures deform \cite{meunier2010diffusive,martinez2018diffusive}.
We note that since the two-phase flow solver is based on a diffuse interface, the fluid velocity will not exactly match the interface velocity.
Thus, the price to pay for avoiding spurious velocity is that particles may, albeit rarely, move from one phase to the other.

To measure Lyapunov exponents, we solve \cref{eq:dxt,eq:drhot} together using an explicit scheme that makes use of local gradients directly available through the finite-element representation (see \textit{SI Appendix}, section B), which our particle-tracking software Partrac uses \cite{linga2025partrac}. 
In particular, we rewrite \cref{eq:drhot} and take the ensemble average to directly obtain the instantaneous stretching rate:
\begin{align}
    \tilde\lambda = \d {\mean{\log \rho}} t = \mean{\hat{\gv \rho} \cdot (\hat{\gv \rho} \cdot \grad) \v u }.
\end{align}
We see that the right hand side only depends on instantaneous quantities, i.e.\ the orientation of infinitesimal filaments, $\hat{\gv \rho}$. Thus we can obtain $\lambda, \lambda_w$, and $\lambda_{nw}$ by using the latter expression and averaging over particles in both phases.
\subsection*{Model for the Dependence of Interface Length on Capillary Number}
\begin{figure}
    \centering
    \begin{tikzpicture}
    \def\tagh{1.6}
    \node [] (A) at (0,-3.8) {\includegraphics[width=0.5\linewidth,clip,trim=0.35cm 0.4cm 0.05cm 0.3cm]{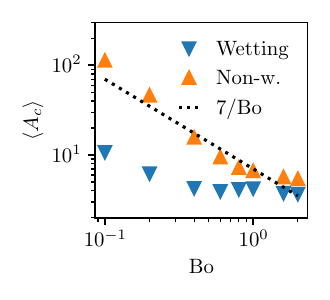}};
    \node [] (Atag) at (-2, -3.8+\tagh) {\textsf{\textbf{A}}};
    \node [] (B) at (4.2,-3.8) {\includegraphics[width=0.5\linewidth,clip,trim=0.35cm 0.4cm 0.05cm 0.3cm]{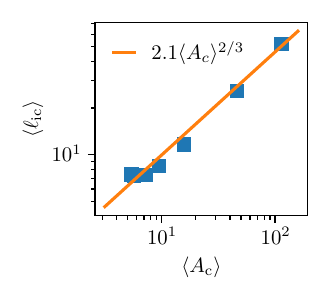}};
    \node [] (Btag) at (2.2, -3.8+\tagh) {\textsf{\textbf{B}}};
    \end{tikzpicture}
    \caption{
    Cluster characteristics of statistically steady 2D two-phase flow simulations in a periodic porous medium.
    (A) Mean cluster size of wetting and non-wetting clusters as a function of $\Bond$.
    (B) Mean interface length for individual clusters $\ell_{\rm ic}$ as a function of mean non-wetting cluster size shows a departure from the expected $\mean{A_{\rm c}}^{1/2}$ scaling.}
    \label{fig:2d_sim_clusters}
\end{figure}
To model how the interface length depends on flow conditions, we use scaling arguments based on the distributions of fluid clusters.
Since the distribution of wetting clusters are dominated by a tail of small clusters where obstacles are nearly in contact, we focus on non-wetting clusters.
We find empirically that the mean non-wetting cluster area $A_c$ scales with Bond number $\Bond$ roughly as:
\begin{align}
    A_c &\sim \Bond^{-1},
\end{align}
as shown in \cref{fig:2d_sim_clusters}A.
This scaling is consistent with \cite{mathiesen2023dynamic}, given that the cluster area scales like $A_c \sim \ell_c^{2}$ where $\ell_c$ is a typical linear cluster size. The number of clusters scales as:
\begin{align}
    N_c &\sim 1/A_c \sim \Bond.
\end{align}
Reminiscent of a fractal structure, we observe that the interface length of individual clusters appears to scale like $\ell_{\rm ic} \sim A_c^{\chi/2}$, with $\chi \simeq 4/3 > 1$ (\cref{fig:2d_sim_clusters}B).
The total interface length then becomes:
\begin{align}
    \ell_{\rm int} & \sim N_c \ell_{\rm ic} \sim A_c^{\chi/2-1} \sim \Bond^{1-\chi/2}.
\end{align}
Combining the latter with \cref{eq:Ca_vs_Bo} gives:
\begin{align}
    \ell_{\rm int} &\sim \Capillary^{3/4-3\chi/8} \sim \Capillary^{1/4},
    \label{eq:ell_int_scaling}
\end{align}
with $\chi \simeq 4/3$.
As shown in \cref{fig:steady2dresults}B, this scaling holds reasonably well up to the point where clusters become large enough to span the system.
Combining \cref{eq:lambda_scaling,eq:ell_int_scaling}, we then end up with the approximate expression \eqref{eq:lambda_combined}.
\subsection*{Experimental Setup}
In the experiments, a transparent quasi-2D porous cell with RSA-placed cylinders was fabricated by stereolithography 3D printing, similar to previous work \cite{vincent2022stable}.
Drainage flows were produced in the cell using air and a 70\% glycerol-water solution by mass, withdrawn from the outlet by a syringe pump.
Detailed methodology of the experimental setup is available in the \textit{SI Appendix}, section C.
The experimental flow rate was $q = \SI{1.0}{\milli\liter\per\minute}$, which in the model geometry entails $\Capillary = 3 \times 10^{-5}$ and $\Peclet = 9\times10^2$, conditions in which the interface motion is marginally intermittent and in the transition region between capillary and viscous fingering \cite{lenormand1988numerical}.
An initial fluorescent line of dyed solution (\SI{20}{\micro\Molar} fluorescein) was produced in the liquid phase by removing a needle from the cell with a translation stage while injecting dye.
The dye was excited with a light-emitting diode panel (wavelength \SI{470}{\nano\meter}) and the fluorescence intensity was imaged through a band-pass filter (520 $\pm$ \SI{5}{\nano\meter}).
A glycerol-water mixture was employed rather than pure water to slow the dynamics for imaging over the required exposure times.
A calibration curve was used to convert imaged fluorescence intensities into solute concentration fields (see \textit{SI Appendix}, section C), producing the time- and space-resolved concentration fields displayed in \cref{fig:experiment}.
\subsection*{Data availability}
All data needed to evaluate the current paper is available in the main text or \emph{SI Appendix}.
}
\showmatmethods{} 
\acknow{
We thank Tom\'as Aquino and Joris Heyman for fruitful discussions.
GL, KP, FR and TLB acknowledge support from the Research Council of Norway (RCN) through their FRINATEK funding scheme, project no.\ 325819 (M4).
GL, KP and MM acknowledge support from the RCN through their Center of Excellence funding scheme, project no.\ 262644 (PoreLab).
GL, JM, FR and TLB acknowledge funding from the Akademia agreement between Equinor and the University of Oslo (Modiflow).
GL and FR thank the Norwegian Centre of Advanced Study for support (CAS project FricFrac to FR; YoungCAS project Mixing by Interfaces to GL). 
Computations were performed on resources provided by Sigma2 - the National Infrastructure for High-Performance Computing and Data Storage in Norway (project NN10091K).
}
\showacknow{} 
%
\bibliography{references}
\end{document}



\maketitle

\SItext

\subsection{Two-phase flow simulations}
\label{sec:tpflow}
\subsubsection{Phase-field model}
To solve two-phase flow in porous domains, we use a phase-field model (specifically, a Navier--Stokes--Cahn--Hilliard model) for matched fluids described by \cite{hohenberg1977theory,jacqmin1999calculation,anderson1998diffuse,ding2007diffuse,abels2012thermodynamically}
\begin{subequations}
\begin{align}
    \pdt \phi + \div (\v u \phi ) &= \div ( M \grad g_\phi ), \\
    g_\phi &= \tilde \sigma (- \epsilon \grad^2 \phi + \epsilon^{-1} W'(\phi)) \\
    \rho ( \pdt{ \v u } + \v u \cdot \grad \v u) - \mu \grad^2 \v u &= - \grad p - \phi \grad g_\phi + \v f \\
    \div \v u &= 0.
\end{align}\label{eq:pfmodel}\end{subequations}
Here, $\phi$ is the phase-field ($\phi \simeq \pm 1$ in the pure phases, and $|\phi| < 1$ in the finite-thickness interface region), $\v u$ is the velocity field, $M$ is a phase-field mobility, which we take to be a constant coefficient, $g_\phi$ a chemical potential for the phase field, $\tilde \sigma = 3 \sigma / (2\sqrt{2})$ where $\sigma$ is the surface tension, $\epsilon$ a numerical interface thickness, $W(\phi) = (1 - \phi^2)^2/4$ a double-well potential, $\rho$ is the density, $\mu$ is the dynamic viscosity, $p$ is a pressure, and $\v f$ is a body force.
At solid boundaries, we apply the following conditions (see, e.g.\ \cite{carlson2012universality}): 
\begin{subequations}
\begin{align}
    \v{u} &= \v{0}, \\
    \hat{\v{n}} \cdot \nabla g_\phi &= {0}, \\
    \hat{\v{n}} \cdot \nabla \phi &= \frac{2\sqrt{2}}{3 \epsilon } \cos (\theta_0) f'_w (\phi) ,
\end{align}\label{eq:pfmodel_bcs}\end{subequations}
where the function $f_w(\phi) \equiv (2 + 3\phi - \phi^3)/4$ smoothly interpolates between 0 and 1 to enforce the difference in the solid-liquid interface energy between the two liquid phases, and $\hat{\v n}$ is the normal vector of the boundary pointing into the solid. We apply periodic boundary conditions to all fields on the horizontal and vertical sides of the domain.
The model \eqref{eq:pfmodel} with \eqref{eq:pfmodel_bcs} is associated with a free-energy functional:
\begin{equation}
    \mathcal{F}[\phi] = \int_\Omega \left[ \frac{1}{2} \tilde \sigma \epsilon |\grad \phi|^2 + \tilde \sigma \epsilon^{-1} W(\phi) \right] \diff V + \sigma \cos \theta_0 \int_\Gamma f_w(\phi) \diff S,
\end{equation}
where $\Omega$ is the fluid domain and $\Gamma$ refers to the solid-fluid boundary. 
The model is thermodynamically consistent in the sense that $\mathcal{F}$ decays in time in the absence of body forces, $\v f = \v 0$ (see, e.g.\ \cite{abels2012thermodynamically}).
To simplify and isolate the main parameters controlling the flow, we divide by the density $\rho$:
\begin{subequations}
\begin{gather}
    \pdt \phi + \div (\v u \phi ) = m \nabla^2 \eta, \quad
    \eta = (- \epsilon \nabla^2 \phi + \epsilon^{-1} W'(\phi)), \\
    \pdt{ \v u } + \v u \cdot \grad \v u - \nu \nabla^2 \v u = - \grad \bar{p} -  \tilde \gamma \phi \grad \eta + F \hat{\v f},
    \quad \div \v u = 0.
\end{gather}\label{eq:pfmodel_reduced}\end{subequations}
Here, $\nu = \mu / \rho$, $\tilde\gamma = \tilde \sigma / \rho$ such that $\gamma = \sigma / \rho $, $\eta = g_\phi / (\gamma \rho)$, $m = M \gamma \rho$, $F = |\v f| / \rho$. We absorb all remaining factors and terms into the reduced pressure $\bar p$.
This shows that the flow properties of the system, such as the mean velocity $U$ [\SI{}{\meter\per\second}], are essentially governed by the microscopic characteristic length scale $d$ [\SI{}{\meter}], the reduced surface tension $\gamma$ [\SI{}{\meter\cubed\per\second\squared}], the kinematic viscosity $\nu$ [\SI{}{\meter\squared\per\second}], the reduced forcing $F$ [\SI{}{\meter\per\second\squared}], the reduced numerical phase field mobility $m$ [\SI{}{\meter\cubed\per\second}] and the numerical phase field thickness $\epsilon$ [\SI{}{\meter}].
These parameters can be combined in the dimensionless groups
\begin{align}
    \textrm{Ca} = \frac{U \nu}{\gamma}, \quad
    \textrm{Bo} = \frac{F d^2}{\gamma}, \quad
    \textrm{Oh} = \frac{\nu}{\sqrt{\gamma d}}, \quad
    \textrm{Sc}_{\textrm{PF}} = \frac{\nu d}{m}, \quad
    \textrm{Cn} = \frac{\epsilon}{d}.
\end{align}
Here, the first three, that is, respectively, the Capillary number, the Bond number, and the Ohnesorge number, reflect the physics of the problem.
The Reynolds number is given by combining these parameters, $\textrm{Re} = U d / \nu = \textrm{Ca} \textrm{Oh}^{-2}$ and is typically small ($<1$) for the flows considered in this work.
The last two numbers, i.e., the phase-field Schmidt number and the Cahn number, are numerical parameters that should ideally be, respectively, as large and small as numerically feasible.

\subsubsection{Numerical scheme}
We solve \eqref{eq:pfmodel_reduced} using finite differences in time and finite elements in space. 
To balance speed, accuracy and robustness, we use a standard, first-order semi-implicit projection method where we reduce the computational complexity by splitting pressure and velocity and solving component-wise where possible.
First, we compute the intermediate velocity $\v u^*$ by solving for each component $u_i^{*}$:
\begin{align}
    \frac{u_i^{*}}{\Delta t} + \v u^{k-1} \cdot \grad u_i^{*} - \nu \nabla^2 u_i^{*} &= - \partial_i p^{k-1} + \frac{u_i^{k-1}}{\Delta t} + \tilde \gamma \phi^{k-1} \partial_i \eta^{k-1} + F_i.
\end{align}
where we have arranged all explicit terms on the right hand side.
We then solve for the pressure increment $\delta p^{*}$:
\begin{align}
    \nabla^2 \delta p^{*} = \frac{1}{\Delta t} \div \v u^{*}, \quad
    p^{k} = p^{k-1} + \delta p^{*}, 
\end{align}
to find $p^k$. 
We then solve for the final velocity $\v u^k$ for each component $u_i^{k}$:
\begin{align}
    u_i^{k} = u_i^{*} - \Delta t \partial_i \delta p^{*}.
\end{align}
To obtain the phase field, we solve the following coupled equation for $(\phi^k, \eta^k)$:
\begin{align}
    \frac{\phi^{k}-\phi^{k-1}}{\Delta t} + \div ( \v u^k \phi^k ) &= m \nabla^2 \eta^k, \\
    \eta^k &= \epsilon^{-1} \overline{W'}(\phi^{k}, \phi^{k-1}) - \epsilon \nabla^2 \phi^k,
\end{align}
where we use the semi-implicit discretization
\begin{equation}
    \overline{W'}(\phi^k, \phi^{k-1}) = W'(\phi^{k-1}) + W''(\phi^{k-1}) (\phi^k - \phi^{k-1}).
\end{equation}
We then proceed to the next time step, $k \leftarrow k+1$.

\subsubsection{Computational details}
The scheme described above was implemented and solved for $P_1$ finite elements using the Twoasis solver \cite{linga2025twoasis}, which is based on the Oasis solver for single-phase flow \cite{mortensen2015oasis} on top of the FEniCS framework \cite{logg2012automated}.
Simulations were run with FEniCS 2019.2.0 on the Sigma2/Fram cluster at NRIS using $4 \times 32$ CPUs and on an internal cluster with 4 Intel® Xeon® Gold 6254 CPUs (3.10 GHz, 72 physical cores).
The Twoasis solver was validated against the benchmark cases in \cite{aland2012benchmark} and the pure two-phase variant of the multiphase electrokinetic solver Bernaise developed by some of the authors \cite{linga2019bernaise}.
We store the velocity and phase fields at equidistant sampling times $t = n \tau$, $n = 0, 1, 2, \ldots $, where $\tau$ is a sampling time step, for use in the Lagrangian particle tracking simulations described below in \cref{sec:lagrangian}. 

\subsubsection{3D two-phase displacement}
This case is implemented as the problem \texttt{Porous3DFlux} in Twoasis.
The numerical parameters used for both the multiphase and single-phase flow simulations are summarized in \cref{tab:parameters_3d_multi}.
The difference between the two simulations is that the multiphase flow simulation is initialized with an interface located at $z = z_0$, i.e.,
\begin{align}
    \phi(\v x, t=0) = \tanh\left( \frac{z - z_0}{\sqrt{2}\epsilon} \right),
\end{align}
whereas the single-phase flow simulation is initialized with $\phi = -1$. In both cases, the inlet conditions are
\begin{align}
    \v u (z=0, t) = u_0 \hat{\v z} \quad \text{and} \quad
    \phi(z=0, t) = -1,
\end{align}
and the outlet conditions are
\begin{align}
    \v u(z=L_z, t) = u_0 \hat{\v z} \quad \text{and} \quad 
    \hat{\v z} \cdot \grad \phi (z=L_z, t) = 0.
\end{align}
Periodic boundary conditions are applied to all fields at the lateral sides.

\begin{table}[htb]
    \centering
    \caption{Parameters for 3D multiphase flow simulations. }
    \begin{tabular}{l c c}
    \hline
         Parameter              & Symbol        & Value \\ \hline
         Density                & $\rho$        & 1.0 \\
         Dynamic viscosity      & $\mu$         & 1.0 \\
         Inlet velocity         & $u_0$         & $10^{-3}$ \\ \hline
         Surface tension        & $\sigma$      & 20.0 \\
         Contact angle          & $\theta_0$    & $\pi/2$ \\
         Phase field mobility   & $M$         & $10^{-4}$ \\
         Interface thickness    & $\epsilon$  & $5 \cdot 10^{-2}$ \\ \hline
         Target bead diameter   & $d$         & 1.0 \\
         Bead diameter          & $d'$          & 0.996 \\
         Transverse dimensions  & $L_x \times L_y $ & $4 \times 4 $ \\
         Domain length          & $L_z$           & 8.0 \\
         Inlet/outlet buffer layer thickness & $\ell_z$ & 0.5 \\
         Initial interface position from bottom & $z_0$ & 1.0 \\
         Porosity (excl.\ buffer layer)   & $\Phi$  & 0.39 \\ \hline
         Time step              & $\Delta t$    & $10^{-2}$\\
         Mesh resolution\tablefootnote{The mesh resolution is computed as $\Delta x \approx \sqrt[3]{V_f/N}$ where $V_f$ is the mesh (i. e., fluid) volume and $N$ is the number of cells.}    & $\Delta x$    & $2.8 \cdot 10^{-2}$\\
         Sampling interval      & $\tau$        & 1.0 \\ \hline
    \end{tabular}
    \label{tab:parameters_3d_multi}
\end{table}

\subsubsection{2D periodic two-phase flow}
We run two-dimensional simulations in periodic porous media using the problem \texttt{Porous2D} in Twoasis.
The physical and geometrical parameters used are listed in \cref{tab:parameters_2d_multi}, and the additional simulation/output parameters for the different cases can be found in \cref{tab:parameters_2d_multi_sims}.

\begin{table}[htb]
    \centering
    \caption{Parameters for periodic 2D multiphase flow simulations.}
    \begin{tabular}{l c c}
        \hline
        Parameter           & Symbol    & Value \\ \hline
        Density             & $\rho$    & 1.0   \\
        Dynamic viscosity   & $\mu$     & 1.0   \\ \hline
        Surface tension     & $\sigma$  & 5.0   \\
        Contact angle       & $\theta_0$& $\pi/3$ \\
        Phase field mobility& $M$       & $10^{-4}$ \\
        Interface thickness & $\epsilon$& $5 \cdot 10^{-2}$ \\
        Saturation          & $s_w$     & 0.5   \\ \hline
        Porosity            & $\Phi$    & 0.62  \\
        Cylinder diameter   & $d$       & 1.0   \\
        Mesh resolution\tablefootnote{The mesh resolution is computed as $\Delta x \approx \sqrt{A_f/N}$ where $A_f$ is the mesh (fluid) surface area and $N$ is the number of cells.} & $\Delta x$ & $2.8 \cdot 10^{-2}$ \\
        Domain size         & $L_x \times L_y$ & $60 \times 90$ \\
        \hline
    \end{tabular}
    \label{tab:parameters_2d_multi}
\end{table}

\begin{table}[htb]
    \centering
    \caption{Simulations time steps and sampling intervals for simulations at different capillary numbers resulting from different driving forces $|\v f|$. The first column refers to the rows in Fig.\ 4 of the main manuscript.}
    \begin{tabular}{c c c c c}
    \hline
        Ref.\ in Fig. 4 & Capillary number $\Capillary$ & Driving force $|\v f|$ & Time step $\Delta t$ & Sampling interval $\tau$ \\ \hline
        & $4.3 \cdot 10^{-4}$ & $5.00 \cdot 10^{-1}$ & $5.00 \cdot 10^{-2}$ & $5.00 \cdot 10^{0}$ \\
        D & $1.1 \cdot 10^{-3}$ & $1.00 \cdot 10^{0}$ & $5.00 \cdot 10^{-2}$ & $2.50 \cdot 10^{0}$ \\
        & $2.6 \cdot 10^{-3}$ & $2.00 \cdot 10^{0}$ & $5.00 \cdot 10^{-2}$ & $1.00 \cdot 10^{0}$ \\
        & $4.3 \cdot 10^{-3}$ & $3.00 \cdot 10^{0}$ & $5.00 \cdot 10^{-2}$ & $1.00 \cdot 10^{0}$ \\
        & $6.2 \cdot 10^{-3}$ & $4.00 \cdot 10^{0}$ & $5.00 \cdot 10^{-2}$ & $5.00 \cdot 10^{-1}$ \\
        C & $8.6 \cdot 10^{-3}$ & $5.00 \cdot 10^{0}$ & $5.00 \cdot 10^{-2}$ & $5.00 \cdot 10^{-1}$ \\
        & $1.7 \cdot 10^{-2}$ & $8.00 \cdot 10^{0}$ & $5.00 \cdot 10^{-2}$ & $5.00 \cdot 10^{-1}$ \\
        B & $2.3 \cdot 10^{-2}$ & $1.00 \cdot 10^{1}$ & $1.00 \cdot 10^{-2}$ & $2.00 \cdot 10^{-1}$ \\
        & $4.8 \cdot 10^{-2}$ & $1.60 \cdot 10^{1}$ & $1.00 \cdot 10^{-2}$ & $2.00 \cdot 10^{-1}$ \\
        & $9.9 \cdot 10^{-2}$ & $3.20 \cdot 10^{1}$ & $1.00 \cdot 10^{-2}$ & $2.00 \cdot 10^{-1}$ \\ \hline 
        A & Single-phase        & $5.00 \cdot 10^{-1}$ & $5.00 \cdot 10^{-2}$ & $5.00 \cdot 10^{-1}$ \\ \hline
    \end{tabular}
    \label{tab:parameters_2d_multi_sims}
\end{table}

\subsection{Lagrangian simulations}
\label{sec:lagrangian}

\subsubsection{Method}
We track material lines and infinitesimal stretchable filaments to quantify stretching in time-dependent numerically-computed flow fields $\v u(\v x, t)$.
In all cases, the basis is to calculate the trajectories $\v x_i(t)$ of passive tracers:
\begin{align}
    \d{\v x_i} t = \v u (\v x_i , t).
    \label{eq:dxdt}
\end{align}
The evolution of material lines and surfaces is described, respectively, in \cite{meunier2010diffusive} and \cite{martinez2018diffusive}.
Considering filaments described by $\delta \gv\ell_i(t) = \v x_i'(t) - \v x_i(t)$, where $\delta \ell_i = |\delta \gv \ell_i| \to 0$ is infinitesimal, we can use \cref{eq:dxdt} to obtain:
\begin{align}
    \d {\delta \gv \ell_i} t = \v u(\v x_i + \delta \gv \ell_i, t ) - \v u(\v x_i, t) = \delta \gv\ell_i \cdot \grad \v u(\v x_i, t) + O(\delta \gv\ell_i^2).
\end{align}
Defining $\gv \rho_i(t) = \delta\gv\ell_i(t)/\delta\ell_i(0)$, stretching of \emph{infinitesimal} filaments is thus given by solving:
\begin{align}
    \d {\gv \rho_i} {t} = \gv \rho_i \cdot \grad \v u (\v x_i, t),
    \label{eq:drhodt}
\end{align}
where $\rho_i = |\gv\rho_i|$ is the elongation of filament $i$.
The velocity field $\v u(\v x, t)$ varies in both space and time and is given by the solution to the Eulerian two-phase flow model described in \cref{sec:tpflow}.

\subsubsection{Numerical implementation}
We solve \cref{eq:dxdt,eq:drhodt} in time using the classical Runge--Kutta discretization scheme (RK4) that is fourth-order accurate in the time-step $\delta t$.
We note that $\delta t < \tau$, i.e., the time step used to sample the Eulerian fields, and typically $\delta t \neq \Delta t$.
To reconstruct the velocity fields $\v u_i$ between the sampling times $t_i$, we use the general linear combination:
\begin{align}
    \v u (\v x, t) = \sum_i \alpha_{i} (t) \v u_i (\v x), 
\end{align}
where $\sum_i \alpha_i (t) = 1$ and linear interpolation can be achieved by choosing the coefficients:
\begin{align}
    \alpha_i(t)= \begin{cases}
    \frac{t - t_{i-1}}{t_{i} - t_{i-1}} &\text{for} \quad t_{i-1} < t < t_{i} \\
    \frac{t_{i+1}-t}{t_{i+1}-t_i}       &\text{for} \quad t_{i} < t < t_{i+1}.
    \end{cases}
\end{align}
To solve \cref{eq:dxdt} from the initial position $\v x^{k}_i$ at a time $t^k$, we compute:
\begin{align}
    \v x^{k+1}_i = \v x^{k}_i + \frac{\delta t}{6} \left( \v k_1 + 2 \v k_2 + 2 \v k_3 + \v k_4 \right), \quad t^{k+1} = t^{k} + \delta t,
    \label{eq:xRK4}
\end{align}
where
\begin{align}
    \v k_1 &= \v u \left( \v x^k_i, t^k \right), \\
    \v k_2 &= \v u \left( \v x^k_i + \frac{\delta t}{2} \v k_1, t^k + \frac{\delta t}{2} \right), \\
    \v k_3 &= \v u \left( \v x^k_i + \frac{\delta t}{2} \v k_2, t^k + \frac{\delta t}{2} \right), \\
    \v k_4 &= \v u \left( \v x^k_i + \delta t \v k_3, t^k + \delta t \right).
\end{align}
To numerically represent $\gv \rho$ without losing precision due to overflow errors, we decompose $\gv \rho = \hat{\gv \rho}^k e^{w}$, where $|\hat{\gv \rho}| = 1$, and track the orientation $\hat{\gv \rho}$ and the logarithm of the elongation, $w$.
To solve \cref{eq:drhodt} from the initial state $\gv \rho^k$, i.e., $(\hat{\gv\rho}^k, w^k)$, we use the same RK4 formula as above,
\begin{align}
    \gv \rho' = \hat{\gv \rho}^k + \frac{\delta t}{6} \left( \v j_1 + 2 \v j_2 + 2 \v j_3 + \v j_4 \right),
\end{align}
where the vector contributions $\v j_j$ stem from the same point evaluations as $\v k_j$, namely:
\begin{align}
    \v j_1 &= \hat{\gv \rho}^k \cdot \grad \v u \left( \v x^k_i, t^k \right), \\
    \v j_2 &= \left( \hat{\gv \rho}^k + \frac{\delta t}{2} \v j_1 \right) \cdot \grad \v u \left( \v x^k_i + \frac{\delta t}{2} \v k_1, t^k + \frac{\delta t}{2} \right), \\
    \v j_3 &= \left( \hat{\gv \rho}^k + \frac{\delta t}{2} \v j_2 \right) \cdot \grad \v u \left( \v x^k_i + \frac{\delta t}{2} \v k_2, t^k + \frac{\delta t}{2} \right), \\
    \v j_4 &= \left( \hat{\gv \rho}^k + \delta t \v j_3 \right) \cdot \grad\v u \left( \v x^k_i + \delta t \v k_3, t^k + \delta t \right).
\end{align}
The updated orientation is given by $\hat{\gv \rho}^{k+1} = \gv \rho' / |\gv \rho'|$ and the logarithm of the elongation is given by $w^{k+1} = w^k + \log |\gv \rho'|$.

\subsubsection{Sheet stretching in 3D flow}
\begin{figure}[htb]
    \centering
    \includegraphics[width=0.6\linewidth]{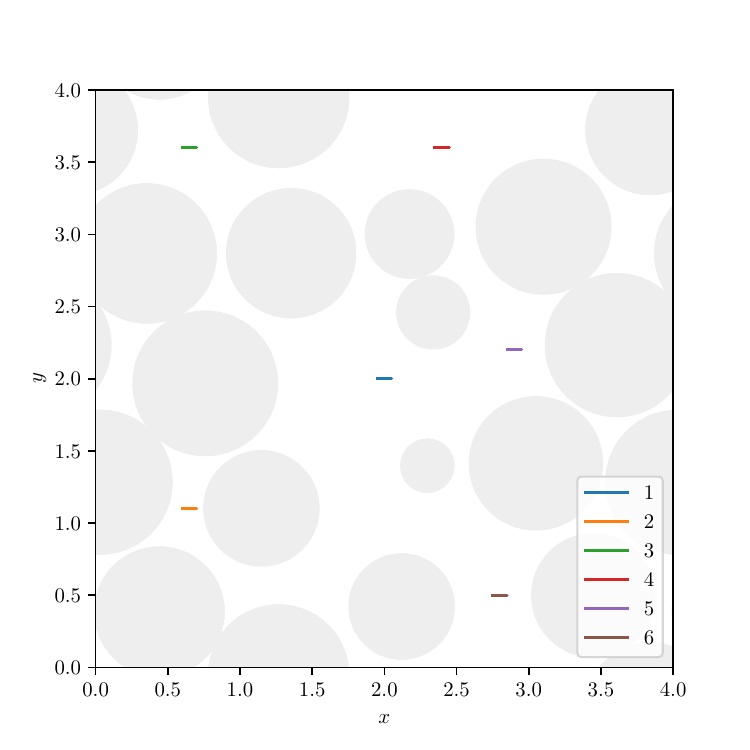}
    \caption{A cut through the $z=1$ plane showing the initial positions of the sheets in the pores.}
    \label{fig:initial_sheets_cut}
\end{figure}

We consider sheets of initial size $0.1 \times 0.1$ which are centered in the plane $z = 1$ and oriented with their normal in the $y$ direction. The plane intersection $z=1$ of the domain is shown in \cref{fig:initial_sheets_cut} alongside the cuts of the initial sheets.
The sheet evolution is tracked by deforming, refining, and coarsening triangles, as described in \citep{martinez2018diffusive}.
Refinement is implemented by successively splitting all edges (and coincident faces) that are longer than $\Delta s_{\rm max} = 0.01$ and introducing a new node at the middle of the edge. Coarsening is performed by collapsing all edges that are shorter than $\Delta s_{\rm min} = 0.002$.
If exactly one of the vertices of the original edge is at the boundary of the sheet, the position of the new vertex is inherited from the vertex at the boundary. Otherwise, the new vertex is placed at the midpoint between the two original vertices.
Sheets are initialized as two triangles spanning the initial square shape, and they are progressively refined until all edge lengths are $\Delta s < \Delta s_{\rm max}$.
For the numerical integration \eqref{eq:xRK4}, we use the time step $\delta t = 0.1$. Refinement and coarsening is performed in time intervals $\tau_{\rm refine} = \tau_{\rm coarsen} = 1.0$.

\subsubsection{Line stretching in 2D periodic flow}
We consider lines that are initially aligned with the $x$-axis, transverse to the flow in the $y$-direction. 
Numerically, we use the diffusive strip method \cite{meunier2010diffusive} where the line segments are described by discrete points that are connected by edges and advected as described in the previous sections. 
Vertices and coincident edges that are initially located inside obstacles are removed and not included in the computations. 
The initial line is represented by line segments spaced by $\Delta s = 0.005$.
When lengths increase beyond $\Delta s_{\rm max} = 0.01$, the edge is refined by splitting it and introducing a new vertex in the middle.
We stop the simulations when the number of vertices representing the line exceeds $10^7$.
The other computational parameters for the four line simulations are shown in \cref{tab:parameters_2D_line}.

\begin{figure}[htb]
    \centering
    \includegraphics[width=0.6\linewidth]{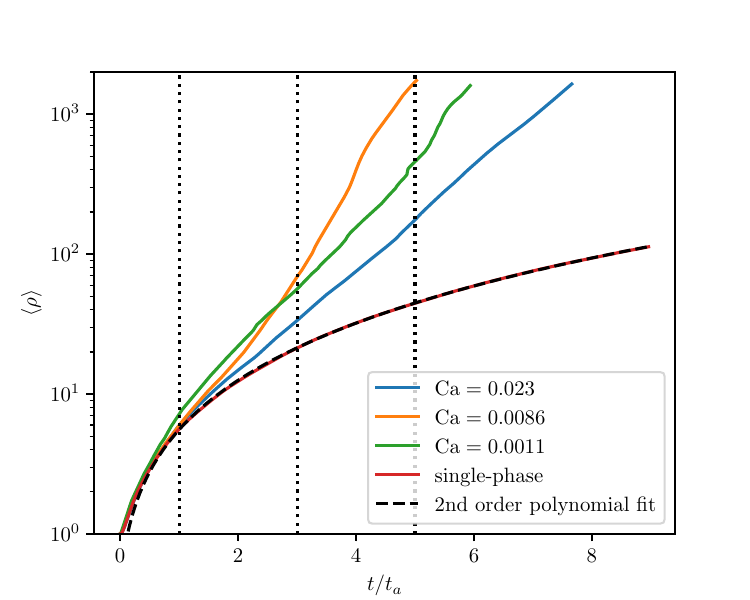}
    \caption{Mean elongation for solute strips in 2D simulations. The vertical dotted lines correspond to the snapshots in Fig.\ 4 in the main manuscript.
    The fastest growth is associated with the optimum $\Capillary \simeq 10^{-2}$ (orange curve) for stretching as measured by the Lyapunov exponent (see Fig.\ 5 in the main manuscript).
    }
    \label{fig:mean_rho_t}
\end{figure}

In \cref{fig:mean_rho_t}, we show the evolution of the mean elongation $\mean{\rho} = \ell(t)/\ell(0)$ of the strips shown in Fig.\ 4 of the main manuscript.
We find that the line length for the single-phase simulation grows algebraically ($\sim t^2$), while for the multiphase flow simulations, it grows exponentially, consistent with the observation of chaotic advection.

\begin{table}[htb]
    \centering
    \caption{Parameters used in the 2D line-stretching simulations for Fig. 4 of the main text.}
    \begin{tabular}{c c c c}
        \hline
        Ref.\ in Fig. 4 & Capillary number $\Capillary$ & Time step $\delta t$ & Refinement interval $\tau_{\rm refine}$ \\ \hline
        B & 0.023           & 0.002 & 0.01 \\
        C & 0.0086          & 0.010 & 0.05 \\
        D & 0.0011          & 0.020 & 0.10 \\
        A & Single-phase    & 0.020 & 0.10 \\ \hline
    \end{tabular}
    \label{tab:parameters_2D_line}
\end{table}

\begin{figure}[htb]
    \centering
    \includegraphics[width=0.6\linewidth]{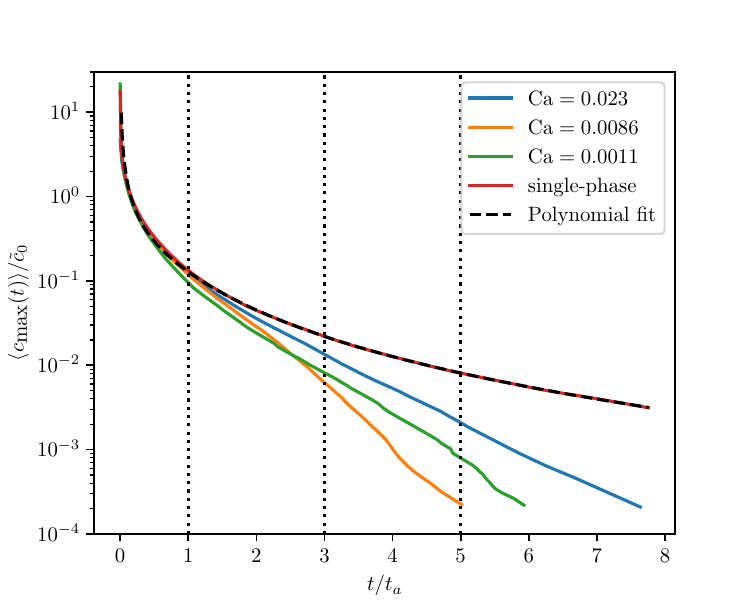}
    \caption{Mean maximum concentration along solute strips in 2D simulations. The vertical dotted lines correspond to the snapshots in Fig.\ 4 in the main manuscript.
    The scaling with $\tilde c_0$ allows us to compare different flow rates at fixed $\Peclet$.
    The fastest decay rate is consistent with the optimum $\Capillary \simeq 0.0086$ for stretching (orange curve). 
    }
    \label{fig:logcmax_mean_t}
\end{figure}

\noindent Using the lamella theory \cite{villermaux2019mixing,heyman2020stretching} we can estimate the scalar decay in the absence of aggregation of line filaments. In particular, at a given location along the strip, the maximum concentration in the direction transverse to the strip is given by
\begin{align}
    c_{\rm max} (t) \approx \frac{c_0}{\sqrt{ 1 + 4 D s_0^{-2} \int_0^t \rho(t')^2 \, \diff t'}},
\end{align}
where $D$ is the diffusion coefficient, and $c_0$ is the initial maximum concentration.
This means that for small initial line widths $s_0$, 
\begin{align}
    \frac{c_{\rm max}(t)}{c_{\rm max} (0)} \approx \frac{s_0}{2 \sqrt{D}} \left( \int_0^t \rho(t')^2 \diff t' \right)^{-1/2}.
\end{align}
Hence, up to a constant pre-factor, $\tilde c_0 = c_0 s_0 / (2 \sqrt{Dt_a}) = c_0 s_0 / (2 d) \Peclet^{1/2}$, $c_{\rm max}(t)$ only depends on the stretching history, i.e.
\begin{align}
    c_{\rm max}(t) \approx \tilde c_0 \left( t_a^{-1} \int_0^t \rho(t')^2 \diff t' \right)^{-1/2}.
\end{align}
We compute the mean of a quantity $\chi$ along the line parametrized by $s$ in the following way:
\begin{align}
    \mean{\chi}(t) = \int_0^{\ell(t)} \chi(s; t) \, \diff s \approx \frac{\sum_i \chi_i(t) \delta \ell_i }{\sum_i \delta \ell_i},
\end{align}
where $\delta \ell_i$ are the discrete filament lengths in the numerical simulation.
The simulation results for the mean of the maximum concentration, $\mean{c_{\rm max}}(t)$, are shown in \cref{fig:logcmax_mean_t}.
The same qualitative impression as shown in \cref{fig:mean_rho_t} persists here; in particular, the single-phase simulations display a slow, algebraic decay, while the multiphase flow simulations exhibit an exponential decay of $\mean{c_{\rm max}}$.
In particular, the multiphase flow conditions that exhibit the fastest stretching are also associated with the fastest scalar decay.
These results illustrate the direct consequences of chaotic advection for mixing.

\subsubsection{Measurements of Lyapunov exponents in 2D periodic flow}
\begin{figure}
    \centering
    \includegraphics[width=0.7\linewidth]{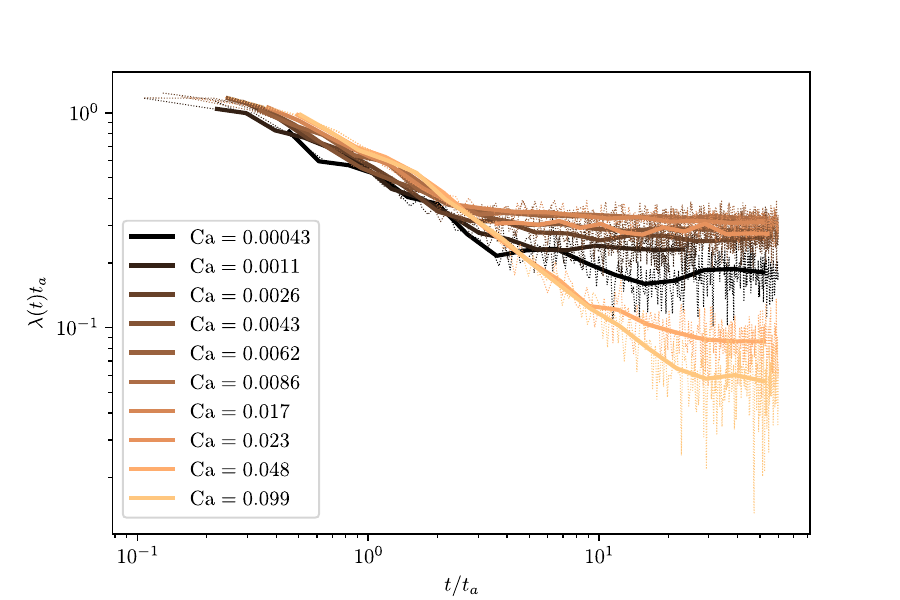}
    \caption{Convergence of Lyapunov exponents in time for different $\Capillary$. The dotted lines represent instantaneous stretching rates based on 10000 filaments. The solid lines represent averages over logarithmically-binned data.}
    \label{fig:lyapunov_convergence}
\end{figure}

The Lyapunov exponent is given by $\lambda = \lim_{t \to \infty} \diff \mean{w_i} (t)/\diff t$ where $\rho_i = \exp(w_i (t))$ and $\mean{\bullet}$ denotes the ensemble mean over filaments that uniformly sample the domain.
Using \cref{eq:drhodt} and the instantaneous orientations $\hat{\gv \rho}_i = \gv \rho_i/\rho_i$, we obtain
\begin{align}
    \d {w_i}{t} = \hat{\gv \rho_i} \cdot ( \hat{\gv \rho_i} \cdot \grad \v u (\v x_i, t) ) \equiv S_i(t).
    \label{eq:drhodt_2}
\end{align}
We note that the right hand side of \cref{eq:drhodt_2} only depends on quantities that are local in time (such as $\hat{\gv\rho}_i$), which allows us to probe the instantaneous stretching rates in the whole system or in the separate phases.
The latter is achieved by constraining the ensemble average to filaments that are located such that the local phase field $\phi(\v x_i) < 0$ or $> 0$.
We run simulations of $10^4$ filaments in all multiphase flow simulations listed in \cref{tab:parameters_2d_multi_sims}, and the Lagrangian time steps are listed in \cref{tab:lagrangian_timesteps}.
The convergence of $\mean{S_i}(t)$ to the Lyapunov exponent $\lambda$ for different capillary numbers $\Capillary$ is shown in \cref{fig:lyapunov_convergence} using filaments in both phases.

\begin{table}[]
    \centering
    \caption{Time steps used for Lagriangian simulations to measure Lyapunov exponents in 2D periodic flow.}
    \begin{tabular}{c c}
\hline
Capillary number $\Capillary$ & Lagrangian time step $\delta t$ \\ \hline
0.00043 & 0.1 \\
0.0011 & 0.02 \\
0.0026 & 0.01 \\
0.0043 & 0.005 \\
0.0062 & 0.005 \\
0.0086 & 0.01 \\
0.017 & 0.002 \\
0.023 & 0.002 \\
0.048 & 0.001 \\
0.099 & 0.0005 \\ \hline
    \end{tabular}
    \label{tab:lagrangian_timesteps}
\end{table}

\subsection{Experimental details}

\begin{figure}[htb]
    \centering
    \includegraphics[width=0.95\linewidth]{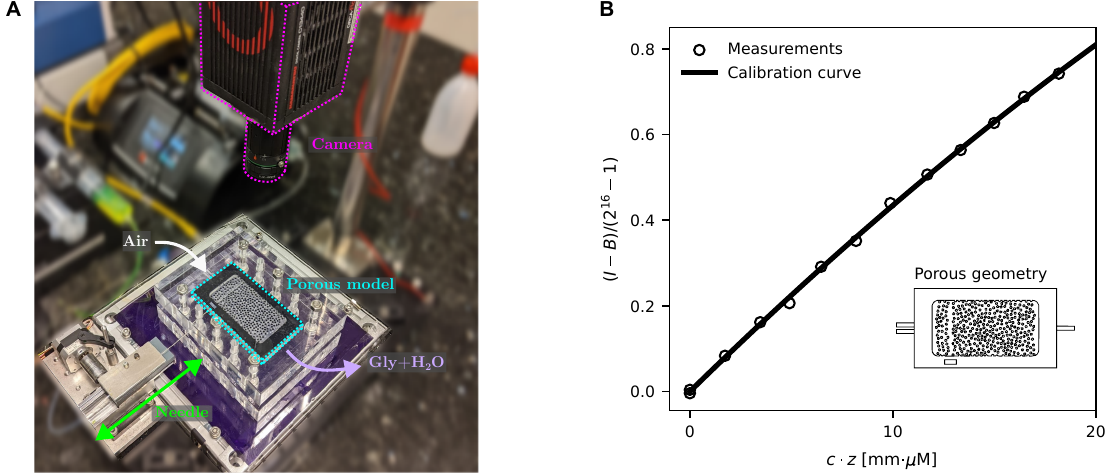}
    \caption{Panel (A) displays the experiment setup, while panel (B) shows the quadratic calibration curve between image intensities and solute concentrations of \Cref{eq:calib}. The line-drawing inset of (B) indicates the RSA arrangement of cylinders in the interior of the porous model.}
    \label{fig:expMeth}
\end{figure}
The millifluidic porous model was designed by placing \SI{2}{\milli\meter} diameter cylinders with random sequential adsorption into a cavity of length 80 mm, width 48 mm, and depth 5 mm using AutoDesk Fusion 360 CAD software.
The design was 3D-printed using a Formlabs 3B SLA printer at resolution \SI{20}{\micro\meter} with Formlabs Clear V4 resin. 
The completed model was sealed between PMMA blocks using thin PVC films as gaskets, and glycerol layers were added between all contacting surfaces to improve model transparency and suppress reflections of the emitted light from solid-air interfaces.
The model was initially saturated with a clear 70\% by mass glycerol-water solution.
An initial line of dye was produced near the model inlet by withdrawing a needle with a translation stage from a side-port in the model while the dyed solution was injected through the needle with a syringe pump.
Once the line was formed, the liquid phase was withdrawn from the outlet port at a flow rate of $q=1.0$ ml/min.
Experimental images were obtained using a Hamamatsu Orca Flash V4 camera at 4MP resolution with a \SI{50}{\milli\meter} f/2.4 lens, flat-image corrected to remove vignetting.
\Cref{fig:expMeth}A highlights the key components of the experimental setup.

The dyed solute was prepared by mixing \SI{20}{\micro\Molar} fluorescein disodium salt in a 70\% by mass glycerol-water solution, made using deionized water and adjusted to $\textrm{pH}=10.0$ with NaOH to maximize fluorescence intensity.
From imaging and optics principles \cite{dahm_measurements_1987}, the flat-corrected images $I$ relate to the depth-averaged concentration field $c$, thickness of the plume $z_p$ along the imaging axis, and the molar absorptivity $\varepsilon$ of the dye via
\begin{equation}
    I = B +  k c z_p e^{-\varepsilon c z_p}.
    \label{eq:calib}
\end{equation}
$B$ is the background fluorescence that occurs in the absence of dye, and the second term represents the emitted fluorescence intensity, with $k$ a constant combining the quantum yield of the dye, the incident light intensity, and the imaging system properties.
For the weak concentrations we consider ($c \approx \SI{30}{\micro\Molar}$) given the thickness ($z_p \approx\SI{0.5}{mm}$) of initial filaments, the second term can be Taylor expanded for a polynomial calibration curve $I-B \approx k c h \left(1-\varepsilon c h\right)$.
We measure $k$ and $\varepsilon$ by injecting a series of initial filaments with different known concentrations and thicknesses, producing the calibration curve displayed in \Cref{fig:expMeth}B.
The inset of this figure indicates the interior geometry of the millifluidic model.
In Fig. 1 of the main text, we scale the concentration field at each time by the maximum concentration on the plume ($\bar{c}_{\rm max}(t)$), estimated as the 99th-percentile of all concentrations on the support mask where $c(\mathbf{x},t)>c_0/100$, $c_0$ being the initial concentration of the dye.

\FloatBarrier

\movie{Experimental resolution of mixing in steady, single-phase flow.}

\movie{Experimental resolution of chaotic mixing in dynamic multiphase flow.}

\movie{Simulation of the stretching of a solute sheet in steady, single-phase flow through a 3D bead pack.
The sheet stretches gently and folds when it occasionally meets stagnation points.
}

\movie{Simulation of the stretching of a solute sheet in dynamic multiphase flow through a 3D bead pack. The sheet stretches and folds when the flow reorients due to interface motion, growing much more in size than its steady single-phase counterpart.}

\movie{Stretching of a solute strip in simulated single-phase flow through a 2D periodic porous medium.
}

\movie{Stretching of a solute strip in simulated multiphase flow through a periodic porous medium at capillary number $\Capillary = 0.023$.}

\movie{Stretching of a solute strip in simulated multiphase flow through a periodic porous medium at capillary number $\Capillary = 0.0086$.}

\movie{Stretching of a solute strip in simulated multiphase flow through a periodic porous medium at capillary number $\Capillary = 0.0011$.}


\bibliography{references}